\magnification 1200
\hsize 15truecm
\vsize 22truecm
\baselineskip 18pt
\null\vskip 1truecm
\nopagenumbers
\centerline {\bf $A_2$ Toda theory in reduced WZNW framework}
\centerline {\bf and the}
\centerline{\bf representations of the $W$ algebra}
\vskip 2.5truecm
\centerline {{\sl Z. Bajnok, L. Palla, G. Tak\'acs}}
\centerline{Institute for Theoretical Physics}
\centerline{Roland E\"otv\"os University}
\centerline{H-1088 Budapest, Puskin u. 5-7, Hungary}
\bigskip
\vskip 3.5truecm
\underbar{Abstract}
\bigskip
Using the reduced WZNW formulation we analyse the classical $W$
orbit content of the space of classical solutions of the $A_2$
Toda theory. We define the quantized Toda field as a periodic primary
field of the $W$ algebra satifying the quantized equations of motion.
 We show that this local operator can be constructed consistently only
in a Hilbert space consisting of the representations corresponding to
the minimal models of the $W$ algebra.
\vfill
\eject

\centerline{\bf 1. Introduction}

The Toda field theories (TT) associated to various simple Lie
algebras ${\cal G}$ have received some interest recently, partly because the
simplest of them, when the Lie algebra is just $A_1$, coincides
with the Liouville theory. It has been known for a long time that
these theories are conformally invariant [1,2] in addition to being
exactly integrable [3].
 Several methods have been suggested
[1,2] to quantize them, and all of these methods showed
convincingly that the quantized versions are bona fide conformal
field theories (CFT). In a recent paper [4] it was shown that tuning
the coupling constant of these theories carefully one can reproduce
the central charges and highest weights of the various \lq minimal'
or \lq coset' models.

 Bilal and Gervais were the first to point out
 that through the Poisson brackets the TT associated to
${\cal G}$ provide a realisation of the $W{\cal G}$ algebras [5].
The concept of $W$ algebras (i.e. extensions of the Virasoro algebra
by higher, (half)integer spin currents) was introduced in the study
of CFT a few years ago [6]. The $W{\cal G}$ algebras
provide a set of $W$ algebras where the spins of the currents ($W_i$),
generating $W{\cal G}$, are determined by the exponents ($h_i$) of
${\cal G}$: $s_i=h_i+1$. Using an essentially free field quantization
it was shown in [2] that the quantized TT provide a systematic
framework to construct the CFT-s that admit the $W{\cal G}$ algebras
as symmetries.

 It has been discovered recently [7] that the classical TT-s can
naturally be viewed as Hamiltonian reductions of the WZNW
theories. This reduction is achieved by imposing certain first class,
conformally invariant constraints on the Kac Moody (KM) currents.
These constraints reduce the chiral KM phase spaces to phase spaces
carrying the chiral $W{\cal G}$ algebras as their Poisson bracket
structures. The advantage of this reduced WZNW description
is that it yields only a restricted set of relevant degrees of
freedom but with a rich algebraic structure as well as giving a
new way to describe the space of classical solutions. A natural
way to quantize these theories is to promote only the relevant
degrees of freedom to operators, trying to preserve
 -- as much as possible -- the boundary
conditions and the algebraic structure.

Recently we carried out this program for the Liouville theory [8]
and -- contrary to our expectation -- we found that this
quantization becomes consistent only in the \lq deep quantum'
domain, but not in the region which is -- at least naively --
smoothly connected to the classical theory.

 The aim of this paper is to show what we can gain both classically
and in the quantum theory from using the WZNW framework to describe
 the $A_2$ TT (which is the next simplest one after
the Liouville theory). In the classical theory we demonstrate that it
enables us to gather information about the \lq classical $WA_2$
algebra  ($W$
for short) representation' (classical $W$ orbit) content of the space
of classical solutions. In particular we shall be able to identify
$W$ orbits that are classical analogues of the quantum highest weight
representations (h.w.r.) both in the singular and in the non singular
sectors of the $A_2$ TT.

In the quantum case we show that promoting only the generators of the
$W$ symmetry and a single Toda field to operators is in a certain
sense a minimal quantization. This means that we require only the
definition of the quantum equivalent of this Toda field be periodic,
and be consistent with $W$ transformation properties, the equation of
motion and locality, but we do not ask for the presence of any
closing operator algebra or any quantum group structure. Yet we show
that when these requirements are supplemented by having a positive
central charge as well as a discrete spectrum of $W$ highest weights
in the Hilbert space ${\cal H}$, where our operators act, then we are
inevitably lead to the conclusion that ${\cal H}$ must consists of
$W$ algebra representations corresponding to the (not
necessarily unitary) minimal models [9], that have no smooth
semiclassical limit. Since the presence of a discrete rather then a
continuos spectrum of $W$ highest weights in ${\cal H}$ corresponds
to the singular sector of the $A_2$ TT we can say that quantizing it
in the reduced WZNW framework works nicely for the singular sector in
the deep quantum domain.

The paper is organized as follows: In section 2. we review the
description of the classical $A_2$ TT in the WZNW framework. Using
this in section 3. we investigate the classical $W$ representation
content of the space of classical solutions. In section 4. we derive
the quantum equation of motion for the Toda field and determine the
general form of the Hilbert space ${\cal H}$, where it may act irreducibly. We
construct the local Toda field and obtain the precise form of ${\cal
H}$ in section 5. We make our conclusions in section 6. The
three appendices, A, B, C,  contain some details about the
way we determined the orbits corresponding to the
 classical highest weights, the way we computed
the various matrix elements of $W_n$ and the way we obtained the
$x\rightarrow x^{-1}$
transformation rule of the generalized hypergeometric functions
respectively.

\centerline{\bf 2. Classical $A_2$ Toda theory in WZNW framework}

  The $A_2$ Toda theory describes the interaction of two
real, periodic scalar fields $\Phi ^a(x^0,x^1)=\Phi ^a(x^0,x^1+2\pi )$;
 $a=1,2$ in two dimensions. Introducing light cone
coordinates $x^{\pm }=(x^0\pm x^1)$ their equations of motion
have the form:
$$\partial _+\partial _-\Phi ^1+2e^{\Phi ^1-{1\over 2}\Phi ^2}=0\eqno(1.1)$$
$$\partial _+\partial _-\Phi ^2+2e^{\Phi ^2-{1\over 2}\Phi ^1}=0\eqno(1.2)$$
The corresponding Lagrangean is
$${\cal L}=\sum\limits_{a,b} {1\over 2}K_{ab}\partial _+\Phi
^a\partial _-\Phi ^b-2\sum\limits_a {\rm exp}({1\over 2}K_{ab}\Phi ^b)$$
where $K_{ab}$ denotes the Cartan matrix of $A_2$. It has been
known for a long time that this theory is conformally
invariant; a property shared by all the other Toda theories (TT).
 The conformal invariance can be seen
from the Feigin Fuchs form of the improved energy momentum tensor:
$$T_{\pm \pm}=\sum\limits_{a,b} {1\over 2}K_{ab}\partial _{\pm }\Phi
^a\partial _{\pm }\Phi ^b-2\sum\limits_a\partial _{\pm }^2\Phi ^a$$
with $T_{+-}=0$. As a consequence $T_{++}=L$ and
$T_{--}=\bar L$  satisfy $\partial _-L=0$, $\partial
_+\bar L=0$ on shell. From ref.[5] we know that in the case of the classical
 $A_2$
TT we have two (commuting) copies of the $WA_2$ algebra ($W$
algebra for short) generated by the spin two $L(x^+)=W_1$ and by
a spin three current $W(x^+)=W_2$, together with their right moving
counterparts $\bar L(x^-)$, $\bar W(x^-)$. (The $W(x^+)$ ($\bar
W(x^-)$) quantities appearing here are somewhat complicated third
order polynomials made of $\partial _+\Phi ^a$, $\partial _+^2\Phi
^a$, $\partial _+^3\Phi ^a$, (resp. $\partial _-\Phi ^a$, $\partial _-^2\Phi
^a$, $\partial _-^3\Phi ^a$) [5,10] but in the following we shall not need
their actual form.)

     Recently a unified description of classical $W$ algebras
associated to the TT-s was
given [7] by exploiting the connection between TT and constrained WZNW
models. In this WZNW
 description the constraints,
 that select the TT in the space of WZNW
currents, generate gauge transformations (left moving upper and right
moving lower triangular Kac Moody (KM) transformations) and the $W{\cal
G}$ algebra is nothing, but the algebra of gauge invariant
polynomials made of the constrained KM current and its
derivatives. One advantage of this approach lies in the fact, that
the brackets between the the $W_i$-s -- which are induced by the
canonical Poisson brackets of the original currents of the WZNW model
-- can be computed readily using some appropriate KM
 transformations that preserve the form of the constrained current [7].
 In the $A_2$ case, when this \lq\lq form"
preserved by the special KM transformation was the \lq\lq highest weight"
one [7] rather than the more familiar \lq\lq Wronskian" one we found that
$$\delta L=\delta W_1=\bigl[a_1(W_1)^{'}+2a_1^{'}W_1-2a_1^{'''}\bigr]+
\bigl[2a_2(W_2)^{'}+3a_2^{'}W_2\bigr]\eqno(2)$$
$$\eqalign{\delta W=\delta W_2
= &\bigl[a_1(W_2)^{'}+3a_1^{'}W_2\bigr]\cr
&+\bigl[a_2\bigl(-{1\over 6}(W_1)^{'''}+{2\over
3}W_1(W_1)^{'}\bigr)+a_2^{'}\bigl(-{3\over 4}(W_1)^{''}+{2\over
3}(W_1)^2\bigr)\cr &-{5\over 4}a_2^{''}(W_1)^{'}-{5\over
6}a_2^{'''}W_1+{1\over 6}a_2^{(V)}\bigr] \cr }\eqno(3)$$
where $a_{1,2}(x^+)$ are the infinitesimal functions characterising
the \lq pure conformal' and \lq pure $W$' parts of the complete $W$
transformations. Eq.(3) shows that $W(x^+)$ transforms as a primary
field of weight $3$ under conformal transformations while its change
under the pure $W$ transformation depends only on the energy momentum
tensor $L=W_1$. Eq.s (2,3) can be converted to the brackets between
the $W_i$-s by
$$\delta W_i=\sum\limits_j\int
dy^1a_j(y)\{ W_i(x),W_j(y) \} \vert _{x^0=y^0}\eqno(4)$$

\noindent In the case of $A_2$ the reduced WZNW framework also
associates to the solutions of the TT an $SL(3)$ valued WZNW field,
$g$, of rather restricted form, containing all the information.
 This approach identifies the fundamental and natural
variables of the $A_2$ TT as the lower right corner element, $u(x^0,x^1)=
u_2(x^0,x^1)={\rm exp}(-{1\over 2}\Phi ^2(x^0,x^1))$, of this
matrix $g$, plus the (chiral) $W_i$ ($\bar W_i$) generators of the
$W$ algebra, since the entire $g$ field can be described in their
terms. The explicit form of $g$ is:
$$g=\pmatrix{\partial _-^2\partial_+^2u+H&\partial
_-^2\partial_+u-{1\over 2}L\partial_+u&\partial _-^2u-{1\over 2}Lu\cr
             \partial _-\partial_+^2u-{1\over 2}\bar L\partial_-u&\partial
_-\partial_+u&\partial _-u\cr
             \partial_+^2u-{1\over 2}\bar Lu&\partial_+u&u}\eqno(5)$$
where $ H=
-{1\over 2}L\partial_+^2u-{1\over 2}\bar L\partial_-^2u+{1\over
4}\bar LLu$. $u_1={\rm exp}(-{1\over 2}\Phi ^1)$ is given as the lower right
subdeterminant of $g$ and this definition is equivalent to eq.(1.2).
On the other hand, ${\rm det}g=1$ -- which is an integral of the ({\sl
linear}) equations of motion for the $u(x^0,x^1)$ field
$$Du=\partial _+^3u-L(x^+)\partial _+u-\bigl(W(x^+)
+{1\over 2}L(x^+)^{'}\bigr)u=0\eqno(6)$$
(plus a similar one, $\bar Du=0$, in the other light cone variable
with $L\rightarrow \bar L$, $W\rightarrow \bar W$) -- implies eq.(1.1).
Regarding $u$, $L$, $W$, and $\bar L$, $\bar W$ as fundamental
variables places the \lq singular Toda solutions' (when $u_1$ and
$u_2$ may have some zeroes) and the \lq non singular' ones (when
$u_1$ and $u_2$ have no zeroes) on an equal footing: both of them are
described by a globally well defined and regular $g$ matrix if
$L(x^+)$, $W(x^+)$ ($\bar L(x^-)$, $\bar W(x^-)$) are non singular,
periodic functions [7]. Using the previously mentioned form preserving KM
transformation to implement the infinitesimal $W$ transformations it
is easy to see that $u(x^0,x^1)$ is a primary field of the $W$
algebra since
$$\eqalign{\delta u=&a_1(x^+)\partial _+u-a_1^{'}u\cr
   &+a_2(x^+)(\partial _+^2u-{2\over 3}L(x^+)u)-{1\over
2}a_2^{'}\partial _+u+{1\over 6}a_2^{''}u\cr }\eqno(7)$$
($u$ transforms in an entirely analogous way under the right moving
algebra generated by $\bar L$, $\bar W$.) If some non singular,
periodic $L$, $W$ ($\bar L$, $\bar W$) are given, then the $u$
field can be constructed from the solutions of the eq.(6) and its
chiral partner as
$$u(x^0,x^1)=\sum\limits_{k=1}^3\psi _k(x^+)\chi _k(x^-)\eqno(8)$$
Here $\psi _k(x^+)$ ($\chi _k(x^-)$) stand for the three linearly
independent solutions of $Du=0$ ($\bar Du=0$) normalized by
$$1=\left|\matrix{\partial ^2\phi _1&\partial ^2\phi _2&\partial ^2\phi _3\cr
                  \partial \phi _1&\partial \phi _2&\partial \phi _3\cr
                  \phi _1&\phi _2&\phi _3\cr }\right| ,\quad \phi
=\psi ,\ \chi ,\quad \partial =\partial _+,\ \partial _-$$

\centerline{\bf 3. Classical representations of the $W$ algebra}

Treating $u(x^0,x^1)$ and the currents of the $W$ algebra as
fundamental variables opens up a new possibility to analyze the space
of classical solutions of $A_2$ TT. As eq.(2,3) and (7) were obtained
from a KM transformation preserving the form of the constrained
current, the transformed quantities,
 $u+\delta u$, $L+\delta L$, $W+\delta W$ will also solve
eq.(6), i.e. the $W$ algebra transforms classical solutions of $A_2$ TT into
another solutions. Therefore the basic object we need is the family
of solutions connected by $W$ transformations: the so called {\sl
orbit} of the $W$ algebra. (In more mathematical terms these $W$
orbits are nothing but the simplectic leafs of the second Gelfand
Dikii bracket [11], which is equivalent to eq.(4) [12].) Clearly these
orbits may be viewed as the classical representations of the $W$
algebra, and to say something about the representation content of the
classical solution space one has to find the invariants
characterizing the orbits. According to a recent study [12] there are just
two types of invariants for the $W$ orbits: a continuous one, the
monodromy matrix $M$, and a discrete one, describing the homotopy classes of
certain non degenerate curves associated to the solutions of $D\psi
=0$. The appearence of the monodromy matrix can be understood in the
following way: though $u(x^0,x^1)$ must be periodic for
$x^1\rightarrow x^1+2\pi $ the solutions of the chiral d.e., (6), may
be quasiperiodic
$$\psi _k(z+2\pi )=M_{kl}\psi _l(z)\eqno(9)$$
if the left and right monodromy matrices are not independent of each
other. Furthermore eq.(7) shows that the $W$ transformations act
linearly on $\psi $ thus they obviously preserve $M$. Of the homotopy
classes it was shown [12] that in the most general case there are just
three of them -- in marked contrast to the Liouville case, when the
 discrete invariant could take infinitely many different values
counting the (conserved) number of zeroes of the Liouville analogue
of the $u$ field [13,14].

Once we can characterize the orbits -- the classical
representations of the $W$ algebra -- the next question is to
determine which of them may correspond to highest weight
representations (h.w.r.). In the (quantum) h.w.r. the
expectation value of the energy operator is bounded below and it
attains its minimum value for the highest weight state, which is
a simultaneous eigenvector of both $L_0$ and $W_0$ [6]. Therefore
it is natural to expect that a $W$ orbit would correspond to a
h.w.r. if the total energy,
 $\int\limits_0^{2\pi }L(z)dz$, stays bounded below as
we move along the orbit. Furthermore we also expect, that it
also contains a solution of eq.(6) (the \lq\lq classical h.w."
vector) with {\sl constant} $L$, $W$, such that
the total energy has at least a local minimum there, i.e.
$\int\limits_0^{2\pi }L(z)dz$ increases if we move away from
this solution along the orbit.

To investigate the representation content of the classical
solution space and in particular to see what parts of it may
correspond to h.w.r. we adopted the following procedure [11]: first
we picked a monodromy matrix $M$, and looked for such $\psi _k$-s
that satisfy eq.(9) and would give constant $L_0$-s and $W_0$-s
through eq.(6). (Technically we determined $L_0$ and $W_0$ using
only two $\psi _k$-s and found the third one from the
normalization condition.) Then, in the second step, by iterating
the transformation leading to eq.(2,3) we determined if $\Delta
L=E(a_1,a_2)-L_0$ -- where $E(a_1,a_2)=\int\limits_0^{2\pi
}L(z)dz$ -- is positive for all (periodic) $a_1$ and $a_2$ or not.
We call a $W$ orbit a potential classical h.w.r. if $\Delta L$
is positive for all $a_{1,2}$ (for details see Appendix A).

So far we analysed only orbits with diagonalizable monodromy
matrices in the generic case, i.e. when all the parameters
appearing are different and nonvanishing. Since $M\in SL(3)$, its
eigenvalues are either all real or it has a complex conjugate
pair of them and a real one. In the former case a large
class of $M$-s can be described by
$$M={\rm diag}\bigl(e^{\Lambda 2\pi },\ e^{m2\pi },\
e^{-(\Lambda +m)2\pi }\bigr)\qquad \Lambda \ne m\eqno(10)$$
where $\Lambda $ and $m$ are arbitrary real parameters. The
$\psi _k(x^+)$ satisfying eq.(9) with this $M$ and yielding the
constant energy and $W$ densities
$$L_0=\Lambda ^2+\Lambda m+m^2\qquad W_0=-m\Lambda (m+\Lambda )\eqno(11)$$
are
$$\psi _1(x^+)=Ne^{\Lambda x^+};\quad \psi
_2(x^+)=Ne^{mx^+};\quad \psi _3(x^+)=Ne^{-(\Lambda +m)x^+};$$
$$N=\bigl[(m-\Lambda )\{ m\Lambda+2(m+\Lambda )^2\}\bigr]^{-1/3}\eqno(12)$$
Since the curve associated to these $\psi _k$-s has at most two
zeroes this solution is in the \lq non oscillatory' homotopy
class in the classification of [12]. It is important to notice,
that $L_0>0$ for all non vanishing $\Lambda $ and $m$. From the
analysis of $E(a_1,a_2)$ around this solution we concluded that
this type of orbits can be classical h.w.r. for {\sl all values} of
$\Lambda $ and $m$. The right moving sector can be obtained from
eq.(11,12) by some trivial substitutions if the monodromy matrix
there has the same form as eq.(10) but with $\Lambda \rightarrow
\hat \Lambda $ and $m\rightarrow \hat m$. Using these chiral
solutions in eq.(8) we see that $u(x^0,x^1)$ will be periodic if
$\hat \Lambda =\Lambda $ and $\hat m=m$, and then
$$u(x^0,x^1)=N\hat N\bigl(e^{2\Lambda x^0}+e^{2mx^0}+e^{-(2\Lambda
+m)x^0}\bigr) $$
i.e. the $A_2$ Toda sector corresponding to these orbits is the
non singular one.

A large class of monodromy matrices having a real eigenvalue as well
as a complex conjugate pair can be described by
$$M(\Lambda ,\rho )=\pmatrix{e^{\Lambda 2\pi }{\rm cos}(\rho \pi )&
                              e^{\Lambda 2\pi }{\rm sin}(\rho \pi )&0\cr
-e^{\Lambda 2\pi }{\rm sin}(\rho \pi )&e^{\Lambda 2\pi }{\rm
cos}(\rho \pi )&0\cr
0&0&e^{-2\Lambda 2\pi }\cr }\eqno(13)$$
where $\Lambda $ and $\rho >0$ are real parameters. We note that
$M(\Lambda ,\rho +2K)=M(\Lambda ,\rho )$, ($K$ integer), thus the
 domain of $\rho $ containing only inequivalent $M$-s is $0<\rho <2$.
Furthermore if $\rho $ is integer ($\not= 0$), then $M$ has three real
eigenvalues (in general a doubly degenerate one a non degenerate one) thus
 some
of these cases correspond to the $\Lambda \rightarrow m$ limit of
eq.(10). We also note, that for $\Lambda =0$, $\rho =2K$, $M$ becomes
the identity matrix.

 The $\psi _k$-s satisfying eq.(9) with this $M$
and yielding constant energy and $W$ densities now have the following
form:
$$\psi _1(x^+)=\tilde Ne^{\Lambda x^+}{\rm sin}{\rho x^+\over
2};\quad \psi _2(x^+)=\tilde Ne^{\Lambda x^+}{\rm cos}{\rho x^+\over
2};\quad \psi _3(x^+)=\tilde Ne^{-2\Lambda x^+}$$
$$\tilde N=\bigl[-\rho ({9\over 2}\Lambda ^2+{\rho ^2\over
8})\bigr]^{-1/3}\eqno(14)$$
while the $L_0$ and $W_0$ densities are
$$L_0=3\Lambda ^2-{\rho ^2\over 4};\qquad W_0=-2\Lambda (\Lambda
^2+{\rho ^2\over 4})\eqno(15)$$
The solution of eq.(6) given by eq.(14,15) is more interesting
 than the one described by eq.(11,12). First of all we note
that now -- unlike in the previous case -- the energy density may be
negative, $L_0<0$, if $\vert \Lambda \vert <\rho /2\sqrt{3}$. In the
$0<\rho <2$ domain the curve associated to this solution is again
in the \lq\lq non oscillatory" homotopy class, but the possibility of
keeping $M$ fixed while shifting $\rho $ by an even integer may
correspond to describing solutions with the same $M$ but belonging to
the \lq higher' homotopy classes. Precisely this happens for $\Lambda
=0$ when $\rho =2,\ 4,\ 6$, since in these cases eq.(14,15) give the three
representative solutions of the three homotopy classes belonging to
$M={\rm Id}$ as discussed in [12].

Analysing the behaviour of $E(a_1,a_2)$ around this solution we
concluded that this type of orbits can be classical h.w.r. for all
values of $\Lambda $ if $\rho <1$. This is surprising since it
implies that the orbit containing the \lq classical $SL_2$ invariant
vacuum' ($\Lambda =0$, $\rho =2$) cannot be a highest weight one.
This result is important as it implies that the quantum theory may
have no smooth semiclassical limit if it contains the $SL_2$
invariant vacuum in a (quantum) highest weight representation.
(One can see in the following way that eq.(14,15) with $\Lambda =0$, $\rho =2$
indeed describe the invariant classical vacuum :
computing the brackets between the $T_n$, $W_n$ Fourier components of
$L(x^+)$ and $W(x^+)$ from eq.(2-4) one finds that the central term
in $\{ T_n,T_m\} $ is of the form ${c\over 12}n^3\delta _{n,-m}$ with
$c=24$. To convert it into the canonical ${c\over 12}n(n^2-1)\delta
_{n,-m}$ form we have to make a shift in $T_0(\equiv L_0)$ by
$c/24=1$, and after this shift the solution with $\Lambda =0$, $\rho =2$
will be the one of vanishing energy and $W$ density. This argument also shows
that all the orbits characterized by $M$-s in the form of eq.(10) will
have an energy density bounded below by $1$.)

The right moving $\chi _k(x^-)$ solutions can again be obtained by
some obvious substitutions from eq.(14,15) if we assume that the right
moving monodromy matrix differs from eq.(13) only in the parameter
replacments $\Lambda \rightarrow \hat \Lambda $, $\rho \rightarrow
\hat \rho $. Using these $\psi _k$-s and $\chi _k$-s in eq.(8) to
construct $u(x^0,x^1)$ we conclude that $u$ will be periodic if $\hat
\Lambda =\Lambda $ and $\rho +\hat \rho =2J$ with $J$ integer. From
the actual form of $u$
$$u(x^0,x^1)=\tilde N \widehat {\tilde N}\bigl(e^{2\Lambda x^0}{\rm
cos}[{\rho -\hat \rho \over 2}x^0+{\rho +\hat \rho \over
2}x^1]+e^{-4\Lambda x^0}\bigr)$$
we see that if $\Lambda \not= 0$ then -- depending on the sign of
$\Lambda $ -- it has zeroes either for $x^0>0$ or for $x^0<0$. This
means that the number of zeroes of $u$ may change in time, but
nevertheless their mere existence implies that this type of orbits
are in the \lq singular sector' of the solution space of the $A_2$
TT.

Clearly for orbits characterized by $M$-s having the form of eq.(13)
 $\rho $ is a
kind of angular variable, thus we expect that in the quantum theory
its eigenvalues would be discrete. Through eq.(15) this would imply
that the Hilbert space corresponding to these orbits contains a
discrete spectrum of $W$ algebra highest weights.

In passing we emphasize that it is a rather special property of the
orbits described so far that they contain representatives (the $\psi
_k(x^+)$-s) yielding constant $L_0$ and $W_0$. When we changed $M$
in eq.(10) slightly
$$M={\rm diag}\bigl(-e^{\Lambda 2\pi },\ -e^{m2\pi },\
e^{-(\Lambda +m)2\pi }\bigr)\qquad \Lambda \ne m$$
we could construct only $\psi _k(x^+)$-s giving periodic and
singularity free $L(x^+)$ and $W(x^+)$ (provided $\vert \Lambda
-m\vert <1$) but we were unable to find  $\psi _k(x^+)$-s giving
constant $L_0$ and $W_0$. The same remark applies to orbits with
monodromy matrices in the form of eq.(10) but belonging to the
higher homotopy classes. Based on these we conjecture that these
orbits would correspond to $W$ representations which are neither
highest nor lowest weight ones. Finally we remark that we did
not inquire the orbits described by non diagonalizable $M$-s the
reason being that the analogous case for the Liouville theory
proved to be rather uninteresting [14].

\centerline{\bf 4. The quantum equation of motion and the representation}
\centerline{\bf space for the Toda field}

Motivated by the succes we gained from using the WZNW framework in
describing the solution space of $A_2$ TT we envisage a quantization
procedure that promotes only the relevant, natural degrees of freedom
$u$, $L$, $W$, $\hat L$, $\hat W$ to operators. This seems to be the
main difference between the earlier approaches [1,2] devoted to quantizing
the ($A_2$)TT and the present one. Certainly our $u$ operator is
equivalent to some of the vertex operators constructed in [2]
applying a modified free field quantization, but our framework
 is different. We are not going to use free fields thus we shall
impose the quantized equation of motion -- whose parameters we
determine from its covariance -- to define our Toda field, while
the equivalent of this equation was verified in
[2] for the particular vertex operator.
 Furthermore we are mainly
interested in quantizing the $A_2$ TT in a domain which
 would correspond to the singular sector of the
classical theory. In our approach we intend to
maintain both the algebraic structure and the boundary conditions
found classically.
 Technically we shall use short distance operator
product expansions (and complexified coordinates)
 which are closer to the spirit of CFT than the
method of canonical quantization.

The Hilbert space where our operators act is a big, reducible
representation of the direct product of the left and right (quantum)
$W$ algebras ${\cal H}={\cal W}_L\otimes {\cal W}_R$. ${\cal W}_L$
(${\cal W}_R$) -- which are supposed to contain h.w.representations
only -- are spanned by the Laurent coefficients of the currents
$L(z)$, $W(z)$:
$$L(z)=W_1(z)=\sum\limits_nL_nz^{-n-2}\qquad
 W(z)=W_2(z)=\sum\limits_nW_nz^{-n-3}\eqno(16)$$
($\bar L_n$, $\bar W_n$ are defined in an analogous way, from now on
we shall give the formulae for the left moving sector only if it can
lead to no confusion.) If $\phi (z,\bar z)$ is any local field from
the operator algebra then the $W^j_n$ ($W^1_n=L_n$, $W^2_n=W_n$)
operators act on it according to [6]
$$W^j_n\phi (z,\bar z)=\oint\limits_z{d\zeta \over 2\pi }(\zeta
-z)^{n+j}W_j(\zeta )\phi (z,\bar z)\eqno(17)$$
$L_n$, $W_n$ satisfy the quantum version of the $W$ algebra [6]:
$$[L_n,L_m]=(n-m)L_{n+m}+{c\over 12}n(n^2-1)\delta _{n+m}$$
$$[L_n,W_m]=(2n-m)W_{n+m}\eqno(18)$$
$$\eqalign{[W_n,W_m]=&{c\over 3\cdot 5!}(n^2-4)(n^2-1)n\delta
_{n+m}+b^2(n-m)\Lambda _{n+m}\cr
&+(n-m)\bigl({1\over 15}(n+m+2)(n+m+3)-{1\over
6}(n+2)(m+2)\bigr)L_{n+m}\cr }$$
where the central charge, $c$, is a free parameter, $\Lambda _n$ is
the composite operator built from the $L_n$-s
$$\Lambda _n=\sum\limits_{k=-\infty }^{+\infty }:L_kL_{n-k}:+{1\over
5}x_nL_n$$
$$x_{2l}=(1+l)(1-l)\qquad x_{2l+1}=(l+2)(1-l)$$
and $b^2=16/(22+5c)$. The algebra given by eq.(18) in terms of
commutators has the same structure as the one obtained from eq.(2-4)
on the level of Poisson brackets of $T_n$ and $W_m$; the only difference
being that some of the constants got changed as a result of
quantization. Indeed from eq.(2-4) we found $b^2_{\rm class}=16/(5c)$
with $c=24$ and $x^{\rm class}_{2l}=x^{\rm class}_{2l+1}=2$, after
rescaling the {\sl classical} $W_n$ by $\sqrt{5/2}$ to guarantee that
the ratio of the central terms in $\{ T_n,T_m\} $ and $\{ W_n,W_m\}
 $ is the same as in eq.(18).

Of the $u(z,\bar z)$ we assume that it is a (periodic) primary field
of the left (and right) $W$ algebra(s):
$$L_nu(z,\bar z)=0\quad n>0\qquad L_0u(z,\bar z)=\Delta u(z,\bar z)$$
$$W_nu(z,\bar z)=0\quad n>0\qquad W_0u(z,\bar z)=\omega u(z,\bar z)\eqno(19)$$
Please note that here $\Delta $ and $\omega $ may differ from their
classical values encoded in eq.(7), but we assume that $u(z,\bar z)$
is a spinless field $\Delta =\bar \Delta $. The crucial assumption
about $u(z,\bar z)$ is that it satisfies the \lq quantized version'
of the equation of motion, eq.(6) (plus its chiral counterpart). This
quantized version differs from the classical one in two respects:
first, since we are dealing with opeators now, all the products
appearing in eq.(6) should be normal ordered, and in addition, as a
result of renormalization, even the coefficients of the various terms
may be different from their classical values. Interpreting the normal
ordered products $:L(z)u(z,\bar z):$, $:W(z)u(z,\bar z):$ etc. as
subtracting the singular terms from the ordinary ones plus using
eq.(19,17) we finally get that $u(z,\bar z)$ should satisfy:
$$\kappa L^3_{-1}u-L_{-2}L_{-1}u-\alpha W_{-3}u-\beta L_{-3}u=0\eqno(20)$$
where the $\kappa $, $\alpha $ and $\beta $ parameters are yet to be
determined. The motivation to assume that the quantization we are
considering keeps the form of the classical equation of motion and
changes only the various coefficients comes from two sources: we saw
that this happened with the defining relations of the $W$ algebra in
eq.(18), and this was found in the case of complete, unrestricted
WZNW theory also in ref.[15].

Eq.(20) clearly has the form of a null vector. The requirement, that
fixes the $\Delta $, $\omega $, $\kappa $, $\alpha $ and $\beta $
parameters, is that this grade $3$ null vector should be {\sl covariant}
 under the $W$ algebra
 i.e. denoting the left hand side of eq.(20) as $\chi $, $\chi $
should be annihilated by all $L_n$, $W_n$, for $n>0$. Because of the
commutation relations, eq.(18), for this it is sufficient if $L_1\chi
=W_1\chi =L_2\chi =0$. Analysing these conditions we found that they
lead to a consistent system of equations for the parameters only if
$u(z,\bar z)$ generates two independent null vectors, one on grade one:
$$2\Delta W_{-1}u-3\omega L_{-1}u=0\eqno(21.1)$$
and one on grade two:
$$AL^2_{-1}u+BL_{-2}u+CW_{-2}u=0\eqno(21.2)$$
where $A/C$ and $B/C$ are somewhat complicated functions of $\Delta $
and $\omega $:
$$A/C=-{3\over 2(1-\Delta)}\bigl[-{\omega \over 4\Delta
}+{\Delta \over 6\omega }\bigr];\
B/C={3\over 2(1-\Delta)}\bigl[\omega -{3\omega \over 2\Delta
}+{\Delta (2\Delta +1)\over 9\omega }\bigr].$$
 The consistency of these two null vectors with eq.(18,19)
(i.e. their {\rm covariance}) determines $\Delta $ and $\omega $ as
functions of $c$. In describing these functions (and the rest of the
parameters) we found it extremely useful to introduce a new real
parameter, $Q$, in place of $c$: $c=2(3-4/Q)(3-4Q)$; then $\Delta $
and $\omega $ become:
$$\Delta ={4Q\over 3}-1\qquad \omega _{\pm}=\pm {\Delta \over
3}\sqrt{2\over 3}\sqrt{5Q-3\over 5-3Q}\eqno(22)$$
This means that for any $Q$ we get two $u$ fields with the same
conformal weight but opposite $\omega $ values; we shall denote by
$u(z,\bar z)$ ($\tilde u(z,\bar z)$) the field with $\omega _+$
(resp. $\omega _-$). Eq.(22) also implies that $u$ is a
$\pmatrix{1&1\cr 1&2\cr }$ field in the classification of ref.[9].
 Using these parameters in the equations
expressing the covariance of eq.(20) we got
$$\kappa =Q^{-1}\quad \alpha _{\pm }=\pm {1\over
\sqrt{6}}\sqrt{(5Q-3)(5-3Q)}\quad \beta ={1\over 2}\bigl({Q\over
 3}+1\bigr)\eqno(23)$$
Since $c$ is invariant under the substitution $Q\rightarrow Q^{-1}$
we get two new solutions from eq.(22,23) by making this change there;
thus the total number of $u$ fields belonging to a fixed $c$ is four.

We can understand the appearence of the fields, $u$ and $\tilde
u$, degenerate in $\Delta $ but having opposite $\omega $-s in
the following way: The algebra described by eq.(18) is left
invariant by the transformation $L_n\rightarrow L_n$,
$W_n\rightarrow -W_n$. Denoting by ${\cal M}$ the
 operator implementing this automorphism: ${\cal M}L_n{\cal
M}^{-1}=L_n$;  ${\cal M}W_n{\cal M}^{-1}=-W_n$, we
find from eq.(19) that
 $L_0{\cal M}u{\cal M}^{-1}=\Delta {\cal M}u{\cal M}^{-1}$;
 $W_0{\cal M}u{\cal M}^{-1}=-\omega {\cal M}u{\cal M}^{-1}$.
Therefore we can write $\tilde u(z,\bar z)={\cal M}u(z,\bar z){\cal
M}^{-1}$ expressing the fact that $u$ and $\tilde u$ provide a
representation of the automorphism. Therefore in the following
we shall treat $u$ and $\tilde u$ on an equal footing.

Looking only at the central charge and the conformal weights of the
solutions described by eq.(22,23) the obvious classical limit
($c\rightarrow \infty $, $\Delta \rightarrow -1$) would be
$Q\rightarrow _-0$. However the whole $Q<0$ ($c>98$) domain is ruled out if we
insist on having {\sl real} $\omega $ and $\alpha $, since this
restricts $Q$ to $3/5\leq Q\leq 5/3$ (which even shrinks to $3/4\leq
Q\leq 4/3$ if we demand $c>0$). Though it may seem surprising that
this entirely {\sl chiral} condition forces us into the \lq deep
quantum' domain, $0<c<2$, it is in fact in accord with the Kac
determinant for the $W$ algebra [16]: from the latter one also finds
that in the $c>98$ domain a $\pmatrix{1&1\cr 1&2\cr }$ field
with a conformal weight given by eq.(22) can be degenerate only for
purely imaginary $\omega $-s. Therefore, with real $\omega $, the
quantization we propose, can be carried out only for $0<c<2$.

The $u(z,\bar z)$ ($\tilde u(z,\bar z)$) operator acting in ${\cal H}$ is
 known if we know
its matrix elements. Since we assumed that ${\cal W}_L$ (${\cal
W}_R$) consist of h.w.r. only it is enough if we know the  matrix
elements of $u$ between highest weight states $\Bigl\vert \matrix{h&\bar h\cr
w&\bar w\cr }\Bigr\rangle $:
$$\matrix{W^j_n\cr \bar W^j_n\cr }\Bigl\vert \matrix{h&\bar h\cr
w&\bar w\cr }\Bigr\rangle =0\qquad n>0\quad j=1,2\eqno(24)$$
$$ \matrix{L_0\cr \bar L_0\cr }\Bigl\vert \matrix{h&\bar h\cr
w&\bar w\cr }\Bigr\rangle =\matrix{h\cr \bar h\cr }\Bigl\vert
\matrix{h&\bar h\cr
w&\bar w\cr }\Bigr\rangle \qquad \quad \matrix{W_0\cr \bar W_0\cr }\Bigl\vert
 \matrix{h&\bar h\cr
w&\bar w\cr }\Bigr\rangle =\matrix{w\cr \bar w\cr }\Bigl\vert
 \matrix{h&\bar h\cr
w&\bar w\cr }\Bigr\rangle $$
{}From conformal symmetry alone it follows that
$$\Bigl\langle \matrix{H&\bar H\cr \Omega &\bar \Omega\cr }\Bigr\vert u(z,\bar
z)\Bigl\vert \matrix{h&\bar h\cr
w&\bar w\cr }\Bigr\rangle =G(H,h,\dots )z^{H-h-\Delta }{\bar z}^{\bar
H-\bar h-\bar \Delta }$$where the constant amplitude, $G$, that
depends on all the parameters characterizing the h.w. states and the
$u$ field is left undetermined. However the equation of motion,
eq.(20), together with the $W_0$ part of eq.(19)
 restrict $G$; indeed sandwiching eq.(20) and the $W_0$ part of eq.(19)
 between h.w.
states and using the freedom to deform the contour in eq.(17) together
with eq.(21,24) we found after a somewhat lengthy computation that
$G$ vanishes unless $y=h+\Delta-H$ and $\Omega $ satisfy (for
details see Appendix B):
$$\eqalign{-\kappa y(y+1)(y+2)+&(y+2)(y+h)-\alpha \bigl\{w+\omega
\bigl[{3\over 2\Delta }y-2\bigl(y(y+1)\beta ^{-1-1}\cr
&+(y+h)\beta ^{-2}\bigr)\bigr]\bigr\}+(\beta -1)(y+2h)=0\cr }\eqno(25)$$
$$\Omega =-w-\omega +\omega \bigl\{{3\over \Delta
}y-\bigl(y(y+1)\beta ^{-1-1}+(y+h)\beta ^{-2}\bigr)\bigr\}\eqno(26)$$
where $\omega \beta ^{-1-1}$ ($\omega \beta ^{-2}$) denote the $A$
($B$) coefficients in eq.(21.2) when $C$ is scaled to $-1$. These
eqations become tractable if instead of $h$ and $w$ characterizing
the h.w. states we introduce two new parameters $a$ and $b$:
$$\eqalign{h(a,b)=&{Q\over 3}(a^2+b^2+ab)-{(Q-1)^2\over Q}\cr
w(a,b)=&{1\over 9}\sqrt{2\over 3}{Q^2(b-a)(2b+a)(2a+b)\over
\sqrt{(5Q-3)(5-3Q)}}\cr }\eqno(27)$$
The vacuum state is described by $a_{\rm vac}=b_{\rm vac}=\pm
(1-Q^{-1})$ while the $a,b$ parameters of the $u$ field are
$a^{(1)}_u=-b^{(2)}_u=1-Q^{-1}$, $b^{(1)}_u=-a^{(2)}_u=2-Q^{-1}$ (the
parameters of the $\tilde u$ field are obtained from these
expressions by interchanging $a$ and $b$). Substituting eq.(27) into
eq.(25,26) one gets that the $u(z,\bar z)$, $\tilde u(z,\bar z)$
fields have nonvanishing transitions only if the $A,B$ parameters of
the final state and the $a,b$ parameters of the initial one are
related as
$$A,\ B= \matrix{b-1,&a+1\cr
                           b+1,&a\cr
                           b,&a-1\cr }\qquad \quad A,\ B=
 \matrix{b+1,&a-1\cr
         b-1,&a\cr
         b,&a+1\cr }\eqno(28)$$
respectively. To understand the meaning of these selection rules
it is important to realise that the scalar product between the
(chiral) h.w. states we are using is $\langle d,c\vert
a,b\rangle \sim \delta_{ac}\delta_{bd}$. Therefore eq.(28) can
be interpreted as saying that $u$ ($\tilde u$) maps the state
$\vert a,b\rangle $ to $\vert a',b'\rangle $ where $a'=B$ and
$b'=A$. This means that by acting on a (h.w.) state with $u$
($\tilde u$) we can shift $a$ and $b$ (in appropriate
combinations) by $\pm 1$. Interestingly, if $a$ and $b$ were
integers characterizing the Dynkin labels of an $SL(3)$ irrep
$[a,b]$, then the $a'$ and $b'$ obtained from eq.(28) in the
case of the $u$ field would have the same form as the Dynkin
labels of irreps appearing in the tensor product $3\otimes [a,b]$
($\bar 3\otimes [a,b]$ for the $\tilde u$ field). This is the
quantum equivalent of the classical property, that the $u$ field
was a specific component of an $SL(3)$ triplet.

{}From the chiral partner of the equation of motion, eq.(20), one
finds an entirely analogous selection rule for the $\bar a,\bar
b$ ($\bar A,\bar B$) parameters characterizing the
transformation properties of the h.w.s. under the right moving
$W$ algebra. A connection between $a,b$ and $\bar a,\bar b$
parameters can be established by requiring $u(z,\bar z)$,
$\tilde u(z,\bar z)$ to be periodic. Indeed looking at the \lq
diagonal' transitions ($\Bigl\langle \matrix{b+1&a\cr
\bar b+1&\bar a\cr }\Bigr\vert u(z,\bar z)\Bigl\vert \matrix{a&b\cr
\bar a&\bar b\cr }\Bigr\rangle $ etc.) one immediately obtaines
that $u$ and $\tilde u$ can be periodic only if
$$\bar a=a-(2N+M)Q^{-1},\qquad \bar b=b+(N-M)Q^{-1}\eqno(29)$$
where $N$ and $M$ are integers. It is also easy to see that the
periodicity of $u$ and $\tilde u$ in the \lq non -- diagonal'
transitions ($\Bigl\langle \matrix{b+1&a\cr
\bar b&\bar a-1\cr }\Bigr\vert u(z,\bar z)\Bigl\vert \matrix{a&b\cr
\bar a&\bar b\cr }\Bigr\rangle $ etc.) together with the
diagonal ones would imply that $Qa$ and $Qb$ are integers.
However, as we shall see later, this possibility is
unacceptable.

Therefore, putting everything together, in the following we
choose the Hilbert space, where our $u$ and $\tilde u$ operators
act irreducibly as
$${\cal H}=\sum\limits_{k,l}{\cal W}_{a_0+k,b_0+l}\otimes \bar {\cal
W}_{a_0+k,b_0+l}\eqno(30)$$
where ${\cal W}_{a_0+k,b_0+l}$ ($\bar {\cal W}$) is the full
left (right) Verma modul corresponding to the h.w.s. $\Bigl\vert
 \matrix{a_0+k&b_0+l\cr
a_0+k&b_0+l\cr }\Bigr\rangle $. (Choosing $N=M=0$ in in eq.(29)
guarantees the absence of \lq non -- diagonal' transitions and
this choice will be forced upon us if --- eventually --- we want
to represent the other two operators -- whose $\Delta $ and
$\omega $ were obtained by the $Q\rightarrow Q^{-1}$
substitution from eq.(22) -- as periodic fields in the same
Hilbert space.) The summation over the integers $k,l$ in eq.(30)
is either infinite or restricted to a subset, but in any case
the Hilbert space, (30), contains at most a discrete infinity of
h.w. modules. We emphasize, that this choice is a very natural
one in view of the selection rules, eq.(28), but is not the only
possibility, since we could start with a Hilbert space
containing a continuum spectrum of $W$ algebra highest weights.
We chose eq.(30) since it naturally corresponds to the set of
singular solutions described in sect.3 and may contain the
$SL_2$ invariant vacuum $\Bigl\vert \matrix{a_{vac}&b_{vac}\cr
a_{vac}&b_{vac}\cr }\Bigr\rangle $.

In the Hilbert space (30) the $u(z,\bar z)$, $\tilde u(z,\bar
z)$ operators are characterized by three types of constant
amplitudes $G_i(a,b)$ ($\tilde G_i(a,b)$) $i=1,..3$:
$$G_1(a,b)=\langle b,a-1\vert u(1,1)\vert a,b\rangle ;\
G_2(a,b)=\langle b+1,a\vert u(1,1)\vert a,b\rangle $$
$$G_3(a,b)=\langle b-1,a+1\vert u(1,1)\vert a,b\rangle ;\
\tilde G_1(a,b)=\langle b,a+1\vert \tilde u(1,1)\vert a,b\rangle \eqno(31)$$
$$\tilde G_2(a,b)=\langle b-1,a\vert \tilde u(1,1)\vert a,b\rangle ;\
\tilde G_3(a,b)=\langle b+1,a-1\vert \tilde u(1,1)\vert a,b\rangle $$
where $\vert a,b\rangle $ is a short notation for $\Bigl\vert \matrix{a&b\cr
a&b\cr }\Bigr\rangle $. The automorphism, ${\cal M}$,
transforming $u$ and $\tilde u$ into each other relates the
constant amplitudes of the $\tilde u$ field to those of $u$:
$$\tilde G_1(a,b)=G_2(b,a);\quad \tilde G_2(a,b)=G_1(b,a);\quad
 \tilde G_3(a,b)=G_3(b,a)\eqno(32)$$
Exploiting the fact that $u(z,\bar z)$ is a real field reduces
further the number of independent constant amplitudes since it implies
$$G_2(a,b)=G_1^*(b+1,a);\qquad G_3^*(a,b)=G_3(b-1,a+1)\eqno(33)$$
{}From eq.(31-33) we see that both the $u(z,\bar z)$ and the $\tilde
u(z,\bar z)$ fields are completely parametrized if we give the
 constant amplitudes $G_1(a,b)$, $G_3(a,b)$ for all $a,b$-s
belonging to ${\cal H}$.

\centerline{\bf 5. Construction of the local Toda fields}

These constant amplitudes will be further restricted by requiring the
$u$, ($\tilde u$) operators to be mutually local. This can be studied by
analysing the behaviour of the 4-point functions; i.e. the
expectation values of the products of two field operators $u(z,\bar
z)$ $u(\zeta ,\bar \zeta )$ ($u(z,\bar
z)$ $\tilde u(\zeta ,\bar \zeta )$) between h.w. states. Conformal
symmetry implies that these 4-point functions have the form:
$$\Bigl\langle \matrix{H&\bar H\cr \Omega &\bar \Omega\cr }\Bigr\vert u(z,\bar
z)u(\zeta ,\bar \zeta )\Bigl\vert \matrix{h&\bar h\cr
w&\bar w\cr }\Bigr\rangle =(z\zeta )^{\lambda }(\bar z\bar \zeta
)^{\bar \lambda }f_{uu}(x,\bar x)$$
where $\lambda ={1\over 2}(H-h)-\Delta $, $x=\zeta /z$, $\bar
x=\bar \zeta /\bar z$. The $f_{\tilde u\tilde u}(x,\bar x)$,
$f_{u\tilde u}(x,\bar x)$ and $f_{\tilde uu}(x,\bar x)$ functions are
defined in an analogous way. The locality of the $u$ ($\tilde u$)
operators requires that the functions describing the expectation
values of the products of identical operators be symmetric under
$x\rightarrow x^{-1}$: $f_{uu}(x,\bar x)=f_{uu}(x^{-1},\bar x^{-1})$ (
$f_{\tilde u\tilde u}(x,\bar x)=f_{\tilde u\tilde u}(x^{-1},\bar x^{-1})$),
 while for
the functions describing the expectation values of the products of
different operators it means that they should go into each other
under $x\rightarrow x^{-1}$: $f_{u\tilde u}(x,\bar x)=
f_{\tilde uu}(x^{-1},\bar x^{-1})$.
 On the
other hand eq.(20) implies that each of the $f(x,\bar x)$ functions
 satisfies an -- in general different -- 3-rd
order linear differential equation in both $x$ and $\bar x$. The
constant amplitudes determine the linear combination coefficients in
the solutions of this d.e. through the boundary conditions at $x=\bar
x=0$ ($z\rightarrow \infty $) where only the h.w. states contribute:
 Indeed inserting a complete system of states between the $uu$
($u\tilde u$) operators and taking the $z\rightarrow \infty$
($x\rightarrow 0$) limit when the descendant states are suppressed we
get schematically:
$$\langle AB\vert u(z,\bar z)u(\zeta, \bar \zeta)\vert ab\rangle
\rightarrow \sum\limits_{c,d}\langle AB\vert u(z,\bar z)\vert
cd\rangle \langle dc\vert u(\zeta, \bar \zeta)\vert ab\rangle
(1+\dots )=$$
$$=(z\bar z\zeta \bar \zeta)^{\lambda }\sum\limits_{c,d}G(AB;cd)G(dc;ab)
(x\bar x)^{h(c,d)-{1\over 2}(h(A,B)+h(a,b))}(1+\dots )\eqno(34)$$
where the summation runs over those highest weight states whose
presence between $\langle AB\vert$ and $\vert ab\rangle $ is allowed
by the selection rules, the dots stand for a polynomial of $x$, $\bar x$
representing the contribution of the descendant states, and
$G(AB;cd)$ ($G(dc;ab)$) denotes the constant ampitude
appropriate for the transition $\vert cd\rangle \rightarrow
\langle AB\vert $ ($\vert ab\rangle \rightarrow
\langle dc\vert $).
Thus the requirement of locality can be translated into a system
of equations for the constant amplitudes. As we shall see this
system, when supplemented by some minor and very natural
additional assumptions, determines them completely.

In the following we first derive the $3$-rd order differential
equations and analyze their general properties then we turn to a
detailed investigation of the various transitions  distinguished
by the number of intermediate states in eq.(34). Because the
automorphism ${\cal M}$ transforms $u$ and $\tilde u$ into each
other there are only two essentially independent $f$ functions:
$f_{uu}(x,\bar x)$ and $f_{u\tilde u}(x,\bar x)$ say. Applying
the same method we described in Appendix B for the three point
function we found that both $f_{uu}(x,\bar x)$ and $f_{u\tilde
u}(x,\bar x)$ satisfy an equation of the form
$$\kappa ({\rm I})-({\rm II})-(\beta -1)({\rm III})-\alpha
 ({\rm IV})=0\eqno(35)$$
where for both functions
$$({\rm I})=(\lambda -2)(\lambda -1)\lambda f-3x(\lambda -2)(\lambda
-1)f^{\prime }+3x^2(\lambda -2)f^{\prime \prime }-x^3f^{\prime \prime
 \prime}\eqno(36)$$
$$\eqalign{({\rm II})=&-2f\bigl({\Delta \over (1-x)^3}+h\bigr)+
\bigl({\Delta \over (1-x)^2}+h\bigr)[\lambda f-xf^{\prime }]\cr
&-{1\over (1-x)^2}\bigl({\lambda \over
x}f+f^{\prime }\bigr)-(x-1)f^{\prime }+\lambda (1+{1\over x})f-x^2(1+{1\over
1-x})f^{\prime \prime }\cr
&+xf^{\prime }[2(\lambda -1)-{1\over 1-x}]+f\lambda [{\lambda \over
1-x}-(\lambda -1)]\cr }\eqno(37)$$
$$({\rm III})=-2f\bigl({\Delta \over (1-x)^3}+h\bigr)
-{1\over (1-x)^2}\bigl({\lambda \over
x}f+f^{\prime }\bigr)-(x-1)f^{\prime }
+\lambda (1+{1\over x})f\eqno(38)$$
and $h=h(a,b)$. The difference between the equations of
$f_{uu}$ and $f_{u\tilde u}$ comes from the matrix element of
$W_{-3}$ appearing in the fourth term of eq.(35): in the case of
$f_{uu}$ it is
$$\eqalign{({\rm IV})=&wf+{\omega \over (1-x)^3}f+{3\omega \over 2\Delta}\bigl[
\lambda f\bigl({x\over 1-x}+{1\over (1-x)^2}\bigr)+
f^{\prime }\bigl({1\over (1-x)^2}-(1-x)\bigr)\bigr] \cr
&+\omega \beta ^{-1-1}\bigl[\lambda (\lambda -1)\bigl({1\over 1-x}-2\bigr)f
+x^2f^{\prime \prime }\bigl({1\over 1-x}-2\bigr)+2\lambda {x\over
1-x}f^{\prime }+\cr &4x(\lambda -1)f^{\prime }\bigr]
+\omega \beta ^{-2}\Bigl[\Delta f{x^2-2(1-x)\over (1-x)^3}+
hf\bigl({1\over 1-x}-2\bigr)+\bigl({1\over 1-x}-\cr &3\bigr)[(x-
 1)f^{\prime }-\lambda (1+{1\over x})f]+\lambda f{2-x^2-2/x\over
 (1-x)^2}+f^{\prime }
{x^3-2(1-x)\over (1-x)^2}\Bigr] \cr}\eqno(39)$$
(here $w=w(a,b)$), while for $f_{u\tilde u}$ we got:
$$\eqalign{({\rm IV})=&wf-{\omega \over (1-x)^3}f-{3\omega \over 2\Delta}\bigl[
\lambda f{3-3x+x^2\over (1-x)^2}+xf^{\prime }{1+x-x^2\over
(1-x)^2}\bigr]\cr
-\omega \beta ^{-1-1}&\bigl[x^2f^{\prime \prime }\bigl({1\over
1-x}+2\bigr) +2xf^{\prime }\bigl({\lambda \over 1-x}-2(\lambda -1)\bigr)+
\lambda (\lambda -1)f{3-2x\over 1-x}\bigr] \cr
-\omega \beta ^{-2}&\Bigl[\Delta f\bigl({1\over (1-x)^3}+{1\over 1-x}\bigr)+
hf\bigl({1\over 1-x}+2\bigr)+xf^{\prime }\bigl({1\over
(1-x)^2}+2\bigr)\cr
&+\lambda f\bigl(-1-{x^2\over (1-x)^2}\bigr)\Bigr] \cr}\eqno(40)$$
(here $\omega =\omega _+$ in eq.(22). Once we obtained the equation
for $f_{u\tilde u}$ for a given transition from eq.(35-38,40) we can
get that of $f_{\tilde uu}$ for the same transition by simply changing
the sign of the $\alpha wf$ term.)

It is no surprise that the differential equations for $f_{uu}$ and
$f_{u\tilde u}$ have three singular points at $x=0$, $x=1$ and
$x=\infty $. First we discuss the properties of the singularity at
$x=1$.  Since $x\rightarrow 1$ corresponds to $z\rightarrow \zeta $
we expect them to contain some information about the short distance
behaviour of the $uu$ ($u\tilde u$) operator products. Therefore they
should depend only on the operators involved but should be
independent of the external states ($\vert ab\rangle $, $\langle
AB\vert $). In the case of $f_{uu}$ from eq.(35-39) we found that the
indices characterizing the solution around $x=1$ ($f_{uu}\sim
(1-x)^{\nu }$) are:
$$\nu _1=1-{4Q\over 3},\quad \nu _2={2Q\over 3},\quad \nu _3=2+{2Q\over 3}
\eqno(41)$$
These $\nu _i$-s imply the appearance of three operators $O_i$ $i=1,..,3$ with
conformal dimensions
$$\Delta _1={4Q\over 3}-1=\Delta,\quad \Delta _2={10Q\over 3}-2,\quad
\Delta _3={10Q\over 3}\eqno(42)$$
in the operator product expansion (OPE) of $uu$.
The appearance of $\Delta $ among the $\Delta _i$-s means that $O_1$
may correspond to either $u$ or $\tilde u$. It is interesting to
observe, that $\Delta _2$ has also the form of $h(a,b)$ in eq.(27)
with $a=\pm (1-Q^{-1})$, $b=\pm (3-Q^{-1})$, thus $O_2$ is a new $W$
primary field propping up in the OPE of
$uu$. On the other hand $\Delta _3$ differs from $\Delta _2$ by a
positive integer indicating that the corresponding operator may be a
($W$) descendant of $O_2$. Repeating the same analysis for the
$f_{u\tilde u}$ function we found that the indices are now given by:
$$\mu _1=2-{8Q\over 3},\quad \mu _2={Q\over 3},\quad \mu _3=1+{Q\over 3}
\eqno(43)$$
These indices imply that the three operators $U_i$, $i=1,..,3$
appearing in the $u\tilde u$ OPE have the following conformal dimensions:
$$\Delta _1=0,\quad \Delta _2=3Q-2,\quad
\Delta _3=3Q-1.\eqno(44)$$
It is natural to assume that $U_1$ is nothing but the identity
operator. $\Delta _2$ can again be written in the form of $h(a,b)$ in
eq.(27) with $a=b=\pm (2-Q^{-1})$, thus $U_2$ is again a new $W$
primary field emerging in the $u\tilde u$ OPE, while $U_3$ can again
be interpreted as a descendant of $U_2$. We note that one pair of
indices is differing by an integer for both $f_{uu}$ and $f_{u\tilde
u}$ and this raises the danger of one of the fundamental solutions at
$x=1$ being logarithmic instead of polynomial [17]. We come back to
this problem soon.

As we mentioned earlier the various transitions defining the various
types of $f_{uu}$ and $f_{u\tilde u}$ functions can be classified
according to the number of intermediate states in eq.(34). In fact we
can use eq.(34) together with the selection rules, eq.(28), to
determine all the non vanishing 4-point functions built on the
initial state $\vert ab\rangle $ and collect the transitions leading
to the same final state $\langle AB\vert $. The six $uu$ transitions
($f_{uu}$ functions) belong to two groups: three of them -- when
$A,B$ are $b,a-2$; $b+2,a$ and $b-2,a+2$ respectively -- have just
one intermediate state, while the other three -- when
$A,B$ are $b,a+1$; $b-1,a$ and $b+1,a-1$ respectively -- have two
intermediate states. Of the seven $u\tilde u$ transitions the
diagonal one, i.e. when $\langle AB\vert =\langle ba\vert $, is a
class of its own by having three intermediate states, while all the
others (with $A,B$ being $b+1,a+1$; $b-2,a+1$; $b-1,a-1$;
 $b-1,a+2$; $b+2,a-1$ and $b+1,a-2$ respectively) have only one
intermediate state.

Our strategy to determine the functions belonging to transitions with
one and two intermediate states is the following: first we analyse
the exponents of $x$ ($\bar x$) appearing in eq.(34) then combining
them with the known indices at $x=1$ (eq.(41,43)) we construct some
trial functions, whose validity we check on the computer using the
symbolic formula manipulating program FORM [18]. Once we completed this
we derive from the requirement of locality the equations for the
constant amplitudes.

In case of the three $f_{uu}$ functions with one intermediate state
(IS) we found that the exponent of $x$ ($\bar x$) in eq.(34) is just
$-Q/3$, i.e. is independent of $a$, $b$. Combining this with the
expression $(1-x)^{2Q/3}$ corresponding to $\nu _2$ in eq.(41) we get
a trial function
$$\bigl(x^{-1}(1-x)^2\bigr)^{Q/3}\bigl(\bar x^{-1}(1-\bar
x)^2\bigr)^{Q/3}\eqno(45) $$
which, in addition to exhibiting the singular solutions at $x=0$ and
$x=1$ is also symmetric under $x\rightarrow x^{-1}$. Using FORM to
substitute this expression into the corresponding equations we
checked that it really solves them. Thus when multiplied by the
appropriate products of $G$-s, eq.(45) yields a complete solution to
the three $f_{uu}$ functions with one IS without any restriction on
the constant amplitudes.

In case of the six $f_{u\tilde u}$ functions with one IS the
exponents of $x$ in eq.(34) ($-Q(1+b-a)/6$, $-Q(1+2a+b)/6$,
$-Q(1-2b-a)/6$, each of them appearing twice) do depend on $a$, $b$.
Furthermore computing the exponents for the ${\bf \tilde uu}$ product
between the \underbar{same} states we found that in each case they differ
from the previous ones as a result of the different IS but only in
replacing $a$ and $b$ by $-a$, $-b$ respectively. Therefore using the
expression $(1-x)^{Q/3}$ corresponding to $\mu _2$ in eq.(43) we get
trial functions
$$\sigma _0(x)\sigma _0(\bar x)(x\bar x)^{-{Q\over 6}(b-a)};\quad
\sigma _0(x)\sigma _0(\bar x)(x\bar x)^{-{Q\over 6}(2a+b)};\quad
\sigma _0(x)\sigma _0(\bar x)(x\bar x)^{{Q\over 6}(2b+a)}\eqno(46)$$
(where $\sigma _0(x)=\bigl(x^{-1}(1-x)^2\bigr)^{Q/6}$)
that again exhibit the correct behaviour at $x=0$ and $x=1$.
Furthermore for these trial functions the $x\rightarrow x^{-1}$
substitution amounts to the replacement $a\rightarrow -a$,
$b\rightarrow -b$. Having checked that these trial functions do solve
the corresponding equations we multiplied them with the appropriate
combinations of constant amplitudes and found the following three
independent equations
$$G_3(a+1,b)G_2(b,a)=G_2(b-1,a+1)G_3(a,b)\eqno(47.a)$$
$$G_3(a,b-1)G_1(b,a)=G_1(b-1,a+1)G_3(a,b)\eqno(47.b)$$
$$G_1(a,b-1)G_1(b,a)=G_1(b,a-1)G_1(a,b)\eqno(47.c)$$
from the requirement of locality, $f_{u\tilde u}(x,\bar x)=
f_{\tilde uu}(x^{-1},\bar x^{-1})$.

In case of the three $uu$ transitions with two IS $f_{uu}$ starts at
$x=0$ as a linear combination of two terms. Motivated by this we
assumed, that at $x=1$ it is also a linear combination of two terms,
namely those, whose singular behaviour is given by $\nu _1$ and $\nu
_2$ in eq.(41). Therefore computing the exponents in eq.(34) we
constructed our trial functions as a sum of terms
$(1-x)^{1-4Q/3}x^{\rm exponent}F(\alpha ,\beta ,\gamma ;x)$ where
 $F(\alpha ,\beta ,\gamma ;x)$ is the usual hypergeometric function,
analytic around $x=0$. To every exponent we determined $\alpha $,
$\beta $ and $\gamma $ from demanding two things: first that the
singularities of the sum at $x=1$ be given by $\nu _1$ and $\nu _2$
and second that the members of the sum be transformed into each
other's  linear combination under $x\rightarrow x^{-1}$. Putting
everything together the trial fuctions for the three $f_{uu}$-s with
two IS can be written in the following compact form:
$$\eqalign{\langle b,a+1\vert uu\vert ab\rangle :\quad &\sigma _1(x)\sigma
_1(\bar x)\bigl[G_3(a,b+1)G_2(a,b)\psi _b(x)\psi _b(\bar x)\cr &+
G_2(a+1,b-1)G_3(a,b)\psi _{-b}(x)\psi _{-b}(\bar x)\bigr]\cr }\eqno(48.a)$$
$$\eqalign{\langle b-1,a\vert uu\vert ab\rangle :\quad &\sigma _1(x)\sigma
_1(\bar x)\bigl[G_1(a+1,b-1)G_3(a,b)\psi _a(x)\psi _a(\bar x)\cr &+
G_3(a-1,b)G_1(a,b)\psi _{-a}(x)\psi _{-a}(\bar x)\bigr]\cr }\eqno(48.b)$$
$$\eqalign{\langle b+1,a-1\vert uu\vert ab\rangle :\quad &\sigma _1(x)\sigma
_1(\bar x)\bigl[G_1(a,b+1)G_2(a,b)\psi _{a+b}(x)\psi _{a+b}(\bar
x)\cr &+
G_2(a-1,b)G_1(a,b)\psi _{-a-b}(x)\psi _{-a-b}(\bar x)\bigr]\cr
}\eqno(48.c)$$
where $\sigma _1(x)=\bigl(x^{-1}(1-x)^2\bigr)^{{1\over 2}-{2Q\over
3}}$ and
$$\psi _b(x)=x^{{1\over 2}[Q(b-1)+1]}F(Q[b-1]+1,1-Q,1+Qb;x).$$
The trick we used to check the validity of these expressions on the
computer was to express the second and third derivatives of the
hypergeometric functions $F$ in terms of $F^{\prime }$ and $F$ using
the hypergeometric differential equation and to verify that the
coefficients of $F^{\prime }$ and $F$ vanish separately in eq.(35-39).

The well known $x\rightarrow x^{-1}$ transformation properties of the
hypergeometric functions (see e.g. [19]) imply that:
$$\psi _b(x)=B_1(b){x^{Q(b-1)+1}\over (-x)^{Q(b-1)+1}}\psi _b(1/
x)+B_2(b){x^{1-Q}\over (-x)^{1-Q}}\psi _{-b}(1/x)\eqno(49)$$
where $B_1(b)={\Gamma (1+Qb)\Gamma (-Qb)\over \Gamma (1-Q)\Gamma
(Q)}$; $B_2(b)={\Gamma (1+Qb)\Gamma (Qb)\over \Gamma (Q[b-1]+1)\Gamma
(Q[b+1])}$. Using eq.(49) in the expressions in eq.(48) to implement
the $x\rightarrow x^{-1}$ symmetry we found that for this
$${G_1(a+1,b-1)G_3(a,b)\over G_1(a,b)G_3(a-1,b)}=\phi (a);\quad
{G_3(a,b+1)G_1^*(b+1,a)\over G_1^*(b,a+1)G_3(a,b)}=\phi (b)\eqno(50.a)$$
$${G_1(a,b+1)G_1^*(b+1,a)\over G_1(a,b)G_1^*(b+1,a-1)}=\phi
(a+b)\eqno(50.b)$$
must hold for the constant amplitudes. Here
$$\phi (b)=-{\Gamma ^2(-Qb)\Gamma (Q[b+1])\Gamma (1+Q[b-1])\over
\Gamma ^2(Qb)\Gamma (Q[1-b])\Gamma (1-Q[b+1])}={s(b+1)\over s(b-1)}
{\Gamma ^2(-Qb)\Gamma ^2(Q[b+1])\over \Gamma ^2(Qb)\Gamma ^2(Q[1-b])}$$
with $s(x)={\rm sin}(\pi Qx)$.

After some straightforward algebra one can show that the solution of
eq.(33), (47) and (50) can be written as:
$$\vert G_3(a,b)\vert ^2=N{\Gamma (Qb)\Gamma (-Q[b-1])\Gamma
(Q[a+1])\Gamma (-Qa)\over \Gamma (1-Qb)\Gamma (1+Q[b-1])\Gamma
(1-Q[a+1])\Gamma (1+Qa)}\eqno(51)$$
$$\vert G_1(a,b)\vert ^2=M{\Gamma (Q[a+b])\Gamma (-Q[a+b-1])\Gamma
(-Q[a-1])\Gamma (Qa)\over \Gamma (1-Q[a+b])\Gamma (1+Q[a+b-1])\Gamma
(1+Q[a-1])\Gamma (1-Qa)}\eqno(52)$$
where $N=N(Q)$ and $M=M(Q)$ are undetermined functions of $Q$. (Very
precisely they still could depend on $a$ and $b$ through such
combinations that stay invariant under $a,b\rightarrow a\pm 1, b\pm 1$.)

The diagonal $u\tilde u$ transition (i.e. when the final state is
$\langle ba\vert $) needs special care since now -- unlike in the
previous cases -- all three singularities at $x=1$ may contribute
raising the danger of a logarithmic singularity. Therefore we
determined the indices of the differential equation one gets from
eq.(35-38,40) with $\lambda =-\Delta $ at $x=0$ and $x=\infty $
first. At the origin we got
$$\nu _1^{(0)}={Q\over 3}(1+b-a);\quad \nu _2^{(0)}={Q\over 3}(1+b+2a);
\quad \nu _3^{(0)}={Q\over 3}(1-2b-a)\eqno(53)$$
which nicely coincide with the exponents computed from eq.(34) -- as
is expected -- while at infinity we found
$$\nu _1^{(\infty )}={Q\over 3}(1+a-b);\quad
\nu _2^{(\infty )}={Q\over 3}(1+2b+a);
\quad \nu _3^{(\infty )}={Q\over 3}(1-b-2a)\eqno(54)$$
Combining the indices in eq.(43), (53) and (54) we see that our
differential equation is of the Fuchs type. At $x=0$ ($x=\infty $)
its solution will be free of logarithms -- thus it may
correspond to our boundary conditions, eq.(34) --
 if none of the index pairs is
differing by an integer [17], i.e. if neither $Qa$ nor $Qb$ is an
integer. If this is the case then we don't have to worry about the
potential logarithmic singularity at $x=1$, since in the lack of an
additional branch point it must be absent. It is also encouraging to
observe that $Qa$, $Qb$ not being integers also guarantees that all
the hypergeometric functions appearing in eq.(48) are indeed well
defined power series.

We solved the differential equation for the diagonal $f_{u\tilde u}$
by realising that factoring out $(1-x)^{Q/3}x^{\nu _i^{(0)}}$
$i=1,..3$ from $f_{u\tilde u}$ in eq.(35-38,40)
 one gets the differential equation
$$\Bigl[x{d\over dx}\prod\limits_{j=1}^2(x{d\over dx}+\beta _j-1)-
x\prod\limits_{k=1}^3(x{d\over dx}+\alpha _k)\Bigr]v=0\eqno(55)$$
satisfied by the generalized hypergeometric function $_3F_2=v$ [19]:
$$_3F_2\Bigl(\matrix{\alpha _1&\alpha _2&\alpha _3\cr
                     \ &\beta _1&\beta _2\cr }\bigl\vert x\Bigr)=
\sum\limits_{n=0}^{\infty }{\alpha _1^{(n)}\alpha _2^{(n)}\alpha _3^{(n)}
\over \beta _1^{(n)}\beta _2^{(n)}}{x^n\over n!}$$
where $\alpha _i^{(n)}={\Gamma (\alpha _i+n)\over \Gamma (\alpha
_i)}$. To every exponent $\nu _i^{(0)}$ we determined the $\alpha
_i$, $\beta _i$ parameters as functions of $a$, $b$ and $Q$ by
matching the coefficients of the various terms we got from the
computer to that of coming from eq.(55). Therefore the complete
diagonal transition has the form:
$$\eqalign{\langle ba\vert u\tilde u&\vert ab\rangle =(z\bar z\zeta \bar \zeta
)^{-\Delta }\sigma _0(x)\sigma _0(\bar x)
\bigl[\vert G_1(a+1,b)\vert ^2I_1(a,b\vert x)I_1(a,b\vert \bar x)\cr &+
\vert G_3(b,a)\vert ^2I_2(a,b\vert x)I_2(a,b\vert \bar x)+
\vert G_1(b,a)\vert ^2I_3(a,b\vert x)I_3(a,b\vert \bar x)
\bigr]\cr }\eqno(56)$$
where
$$I_1(a,b\vert x)=x^{Q({1\over 2}+{1\over 3}[2a+b])}\
_3F_2\Bigl(\matrix{Q&Q(1+a+b)&Q(1+a)\cr
                     \ &1+Q(a+b)&1+Qa\cr }\bigl\vert x\Bigr)\eqno(57.a)$$
$$I_2(a,b\vert x)=x^{Q({1\over 2}+{1\over 3}[b-a])}\
_3F_2\Bigl(\matrix{Q&Q(1+b)&Q(1-a)\cr
                     \ &1+Qb&1-Qa\cr }\bigl\vert x\Bigr)\eqno(57.b)$$
$$I_3(a,b\vert x)=x^{Q({1\over 2}-{1\over 3}[2b+a])}\
_3F_2\Bigl(\matrix{Q&Q(1-a-b)&Q(1-b)\cr
                     \ &1-Q(a+b)&1-Qb\cr }\bigl\vert x\Bigr).\eqno(57.c)$$
(The diagonal $f_{\tilde uu}$ function between the same states can be
obtained from eq.(56), (57) by replacing $a$ and $b$.) In Appendix C
we derive the $x\rightarrow x^{-1}$ transformation rule for the
generalized hypergeometric functions $_3F_2$; using them and the form
of the constant amplitudes given in eq.(51), (52) after a somewhat
lenghty calculation we found that $f_{u\tilde u}(x,\bar x)=
f_{\tilde uu}(x^{-1},\bar x^{-1})$ is satisfied for the diagonal
transition if $N(Q)$ and $M(Q)$ appearing in $G_3(a,b)$ (resp.
$G_1(a,b)$) are equal $M(Q)=N(Q)$, and are indeed independent of $a$,
$b$.

 Clearly to determine the actual $Q$ dependence of $M$ we have to
impose some sort of normalization in addition to $u(z,\bar z)$,
$\tilde u(z,\bar z)$ being local operators. We may require that the
dimension zero operator appearing at $x=1$ in the $u\tilde u$ product
be the true identity operator, or, equivalently, that the operator
with conformal dimension $\Delta $ emerging at $x=1$ in the $uu$
product be the correctly normalized $u$ or $\tilde u$. We chose the
technically simpler second possibility. Comparing the initial and
final states in eq.(48) to the selection rules it is clear that $O_1$
can be identified with $\tilde u$ and requiring that the residues of
the $(1-x)^{\nu _1}$ singularity in the three expressions in eq.(48)
be the properly normalized matrix elements of $\tilde u$ we found that
$$M(Q)=N(Q)=\Bigl({\Gamma (Q)\Gamma (2-2Q)\over
 \Gamma (1-Q)\Gamma (2Q-1)}\Bigr)^2\eqno(58)$$
Thus we see that requiring $u$ and $\tilde u$ to be local operators
together with this normalization condition indeed completely
determines the constant amplitudes. We may think of the
identifications $U_1\sim {\rm Id}$, $O_1\sim \tilde u$ as the quantum
equivalents of the classical conditions ${\rm det}g=1$ and ${\rm exp}(-
{1\over 2}\Phi ^1)$ being the lower right subdeterminant of $g$
respectively: just as in the classical case they are automatically
satisfied as a consequence of the equations of motion, apart from an
overall normalization. With this remark we end the constructuion of
the quantized Toda fields $u$ and $\tilde u$, and in the following we
analyze the properties of this solution.

In the first step we rewrite $\vert G_1\vert ^2$ and $\vert G_3\vert
^2$ in a form more suitable for our purposes:
$$\vert G_3(a,b)\vert ^2=N\pi ^4S_1(Q,a,b)\bigl(\Gamma (Qb)\Gamma
(-Q[b-1])\Gamma
(Q[a+1])\Gamma (-Qa)\bigr)^2$$
$$\vert G_1(a,b)\vert ^2=M\pi ^4S_2(Q,a,b)\bigl(\Gamma (Q[a+b])\Gamma
(-Q[a+b-1])\Gamma
(-Q[a-1])\Gamma (Qa)\bigr)^2\eqno(59)$$
where
$$S_1(Q,a,b)=s(Qb)s(Q[b-1])s(Qa)s(Q[a+1])$$
$$S_2(Q,a,b)=s(Q[a+b])s(Q[a+b-1])s(Qa)s(Q[a-1])\eqno(60)$$
The expressions on the left hand side of eq.(59) should be non
negative by definition. However, because of the sine factors, the
expressions on the right hand side may change sign as $a$ and $b$ run
through their domain in eq.(30). Of course our construction of the
(local) $u$ and $\tilde u$ operators makes sense only if this does
not happen; i.e. if for no $a,b$ belonging to ${\cal H}$ is either
$S_1$ or $S_2$ negative. So our remaining task is to find out the
values of $Q$ and the domain of $a,b$ guaranteeing this. We emphasize
that the condition that the modulus squared of a complex number be
non negative has nothing to do with the possible (non)unitarity of
the $W$ representation built on the h.w. state $\vert ab\rangle $.

Looking at eq.(59,60) we note that in the case of
\underbar{irrational} $Q$-s starting from a state $\vert
a_0b_0\rangle $ (with $Qa_0\ne $integer, $Qb_0\ne $integer) we can
never \lq stop' again, i.e. applying $u$ and $\tilde u$ sufficiently
many times to $\vert
a_0b_0\rangle $ we can change the $a,b$ parameters of the final state
to differ from $a_0$ and $b_0$ by any integer without ever finding a
vanishing $G_1$ or $G_3$. This clearly poses a problem since then the
sine factors in eq.(59,60) will sooner or later change sign
contradicting the positivity of $\vert G_1\vert ^2$ and $\vert
G_3\vert ^2$.

If $Q$ is \underbar{rational}; $Q=r/s$ with $r,s>0$ coprime integers,
 then it is
conceivable that starting from a h.w. state $\vert
a_0b_0\rangle $, after applying several times $u$ and $\tilde u$, we
arrive at a final state for which some of the constant amplitudes
vanish; i.e. in this case -- at least in principle -- we may be able to \lq
stop'. However this possibility raises the danger of having $a$-s and
$b$-s in ${\cal H}$ with the unacceptable property $Qa=$integer,
 $Qb=$integer. (In addition in this case we also have to worry about
some of the $\Gamma$ function's arguments becoming a negative
integer.) We may resolve this problem if we can find a domain in the
$(a,b)$ plane such that the constant amplitudes that would correspond
to transitions leading out of the domain vanish on its border, but
inside (or on the border) there are no points for which $Qa$ or $Qb$
is integer. This may happen as the the six possible transitions from $\vert
ab\rangle $ listed in eq.(28) are characterized by different values
of $G_1$ and $G_3$. Pictorially they can be represented as on Fig.1.
 (This means e.g. that the transition keeping $b$ fixed while
increasing $a$ by $1$ is characterized by $G_1^*(a+1,b)$.)
%***  Insertion of Figure 1. ***
\topinsert
\def\mrs#1{\smash{
                 \mathop{\longrightarrow}\limits^{#1}
                }
         }

\def\mdd#1{\Big\Downarrow\
              \rlap{$\vcenter{\hbox{$\scriptstyle#1$}}$}
         }
\def\mus#1{\Big\uparrow\
              \rlap{$\vcenter{\hbox{$\scriptstyle#1$}}$}
         }
\def\mud#1{\Big\Uparrow\
              \rlap{$\vcenter{\hbox{$\scriptstyle#1$}}$}
         }
\def\mls#1{\smash{
                 \mathop{\longleftarrow}\limits^{#1}
                }
         }
\def\mnws#1{\nwarrow\ 
              \rlap{$\vcenter{\hbox{$\scriptstyle#1$}}$}
         }
\def\mses#1{\searrow\
              \rlap{$\vcenter{\hbox{$\scriptstyle#1$}}$}
         }
$$
\matrix{
 & & & & G_3(b,a) & & G_1^*(b+1,a)
       & & & 
\cr \noalign{\medskip}
 & & & & & \mnws{} & \mud{}
 & & &
\cr \noalign{\medskip}
 & & & & G_1(a,b) & \mls{} & (a,b)
                  & \mrs{} & G_1^*(a+1,b) &
\cr \noalign{\medskip}
\mus{b} &  & & & & & \mdd{} & \mses{} & &
\cr \noalign{\medskip}
 \cdot & \mrs{a}
   & & & & & G_1(b,a) & & G_3(a,b) &  \cr} 
$$
\vskip 2mm
\baselineskip=10pt
\leftskip=1cm
\rightskip=1cm
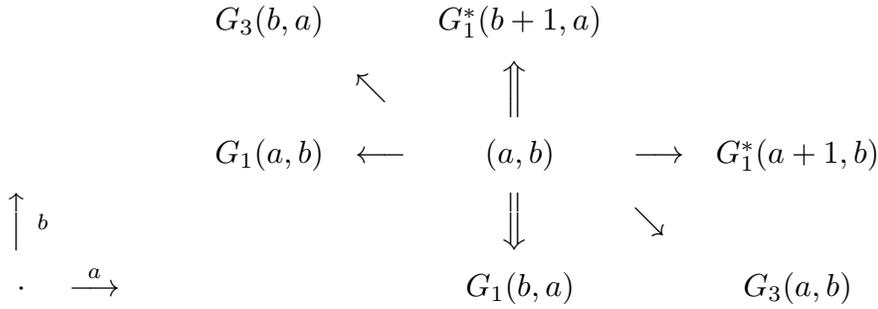
\centerline{Fig.1}
\vskip 5mm
%\leftskip=0cm
%\rightskip=0cm
%\baselineskip=16pt
\endinsert

The domains where all of our conditions are met are triangular ones
with two sides being paralel to the $a$ and $b$ axis
($a\equiv a_0=1-L(s/r)$; $b\equiv b_0=1-K(s/r)$ where $K,L\ge 1$
are integers also
satisfying $K+L\le r-1$) and the third one inclining at $135^{\circ
}$ to the positive $a$ axis ($a+b+1=s(r-K-L)/r$). These domains are
characterized by the two positive integers $K,L$ with $K+L\le r-1$,
and the set of $a$-s and $b$-s belonging to the domain have the form
$$a=1+l-L{s\over r};\qquad b=1+k-K{s\over r};\qquad 1+l+1+k\le s-1
\eqno(61)$$
where $l$ and $k$ are non negative integers. This means that for each
$Q=r/s$ and $K,L$ we get a Hilbert space ${\cal H}_{KL}$ where the
 local $u$ and
$\tilde u$ operators act irreducibly and are defined consistently if
we take the sum in eq.(30) to run over the $a,b$-s in eq.(61).
${\cal H}_{11}$ is the Hilbert space containing the $SL_2$
invariant vacuum with $a_{\rm vac}=b_{\rm vac}=1-(s/r)$. The \lq
largest' Hilbert space where $u$ and $\tilde u$ are defined
consistently is the union of the irreducible ${\cal H}_{KL}$-s:
${\cal H}=\sum\limits_1^{K+L\le r-1}\oplus {\cal H}_{KL}$. Using
eq.(61) and (27) it is easy to see that ${\cal H}$ consists of
nothing but the $W$ representations characterising the (not
necessarily unitary) minimal models [9] belonging to $c(Q=r/s)$.
We also remark that the set of $a,b$-s in eq.(61) and the
maximal ${\cal H}$ are identical to the ones we obtain if we
quantize the Toda fields with the other ($Q\rightarrow Q^{-1}$)
choice for $\Delta $ and $\omega $. More precisely using the
$a$-s and $b$-s one gets from eq.(61) by keeping $l$ and $k$
fixed while leting $K$ and $L$ run in $K,L\ge 1$; $K+L\le r-1$,
from eq.(30) we obtain a Hilbert space ${\cal H}_{kl}$ providing a
representation for the other two Toda fields. The whole ${\cal
H}$ is obtained if we insist on the simultaneous presence of
both types of Toda fields.

\centerline{\bf 6. Conclusions}

In this paper we investigated the $A_2$ TT describing it in the
reduced WZNW framework. In the classical theory working out this
framework in the less familiar \lq highest weight gauge' [7] we
identified the relevant variables as a single Toda field,
$u(z,\bar z)$ and the generators of the classical $W$ symmetry.
Using them we showed that the space of classical solutions can
be divided into classical representations of the $W$ algebra,
the $W$ orbits, that are characterized by the monodromy matrix
and a discrete invariant. We determined two types of monodromy
matrices guaranteeing that the orbits belonging to them are of
the classical highest weight type, in addition to lying in the
singular and non singular sectors of the $A_2$ TT respectively.
Surprisingly, we found that the orbit corresponding to the classical
$SL_2$ invariant vacuum is not of the highest weight type.

In the quantum theory we promoted only the Toda field $u(z,\bar z)$ and the
generators of symmetries to operators. Working in a Hilbert
space containing only at most a discrete infinity of $W$ highest
weight representations we defined $u(z,\bar z)$ as a periodic
primary field satisfying the quantized equation of motion. We
constructed this $u(z,\bar z)$ operator -- and its partner,
$\tilde u(z,\bar z)$, generated from it by the automorhism of
the algebra -- in two steps: first by deriving the selection
rules we determined the types of constant amplitudes
parametrising them, then by imposing their locality through the
4-point functions we determined these constant amplitudes completely.
As a result we learned that these local Toda fields can be defined
consistently if the $Q$ parameter determining the central charge
as $c(Q)=2(3-4/Q)(3-4Q)$ is rational and the Hilbert space is
the collection of $W$ representations corresponding to the
minimal models. We find these results interesting as we arrived
at them without ever demanding the presence of a closing
operator algebra or any quantum group structure.

Summarizing we can say that the reduced WZNW framework gave new
insights both in the classical and in the quantum versions of the
$A_2$ TT. In the quantum case we also see that to go beyond the
minimal models we have to drop some of our assumptions. The obvious
possibilities are to replace the assumption about the representation
content of the Hilbert space by something else and/or to drop one of
the basic axioms of CFT, namely the equivalence between states and
fields, that underlined our computations.

\centerline{\bf Acknowledgements}

One of us (L.P.) thanks J. Balog, P. Forg\'acs and L. Feh\'er for
illuminating discussions at various stages of this investigation.

\centerline { \bf Appendix A }

\centerline { Search for orbits of the classical highest weight type.}

We argued in sect.3 that the classical analogues of the quantum highest
weight states are solutions of eq.(6) with constant $L$ and $W$ such
 that the value
of $\int\limits_0^{2\pi }L(z)dz$ is bounded below
along the orbit. Thus the classical highest weight state must correspond
to at least a local minimum of this integral.

Let's define
$$ l=\int\limits_0^{2\pi }L(z)dz,\qquad  w=\int\limits_0^{2\pi }W(z)dz $$
If $L$ and $W$  are constant (resp. $L_0$, $W_0$) then
the transformation rules of the classical $WA_2$ algebra ( eq.(2-3))
simplify
$$\delta L=[2a_1^{,}L_0-2a_1^{,,,}]+3a_2^{,}W_0\eqno(A.1) $$
$$\delta W=3a_1^{,}W_0+[{2\over 3}a_2^{,}W_0^2-{5\over 6}a_2^{,,,}W_0
+{1\over 6}a_2^{(V)}]\eqno(A.2) $$
Using these to compute the changes in $l$ and $w$ we see that
$\delta l=0 $ as well as $\delta w=0 $
( since $a_1$
and $a_2$ are periodic ) i.e. the points of constant $L$ and $W$ are
stationary points of $l$ and $w$ along the orbit.

Being a classical highest weight state requires also
$$ \delta \delta l\geq 0\eqno(A.3) $$
We call this the stability condition.

We can calculate the concrete formula for $\delta \delta l$
by iterating the $WA_2$ transformation laws ( in the second step we
have to use the full eq.(2-3) as after the first step $L$ and $W$ are
no longer constants ). Discarding total derivative terms we find
$$ \delta \delta l=\int\limits_0^{2\pi }(a_1^{,}\delta
L+a_2^{,}\delta W)dz $$
which can be rewritten as
$$ \delta \delta l=\int\limits_0^{2\pi }
   \pmatrix { a_1^{,} & a_2^{,} }
   \pmatrix { 2L_0-2{d^2\over dz^2 } & 3W_0 \cr
    3W_0 & {2\over 3}L_0^2-{5\over 6}{d^2\over dz^2}+{1\over6}{d^4\over dz^4}
   \cr }
   \pmatrix { a_1^{,} \cr a_2^{,} }dz \eqno(A.4) $$
This is a quadratic form in terms of $a_1^{,}$ and $a_2^{,}$ and the
stability condition amounts to its positive definity. We take an orthogonal
basis in the space of ( $a_1^{,}$,$a_2^{,}$ )  of the form
$$ \pmatrix { a_1^{,} \cr a_2^{,} }={\bf q}e^{inz} ,\qquad n\not= 0 $$
In the subspace of given $n$ the matrix appearing in eq.(A.4)
takes the form
$$ M(L_0,W_0)= \pmatrix { 2L_0+2n^2 & 3W_0 \cr
    3W_0 & {2\over 3}L_0^2+{5\over 6}n^2+{1\over6}n^4
   \cr } $$
The positive definity means that the eigenvalues of this matrix
$$ \lambda _{1,2} (L_0,W_0,n)=a+b\pm \sqrt { (a+b)^2-4ab+9W_0^2
}, $$ where $a$ and $b$ are
$$ a=L_0+n^2 ,\quad b={1\over 12} \bigl[ (2L_0+n^2)^2+L_0n^2 \bigr]
, $$ must be positive for all $n$.
Consequently
$$ a+b>0 ,\quad 4ab>9W_0^2 \quad {\rm for \  all }\quad \mid n \mid \geq 1 $$
The first of these conditions is satisfied iff $ L_0 > -1 $.The
second one
is satisfied for all values of $n$ iff it holds for $n=1$. Therefore
we have the following inequalities for stability
$$ L_0>-1 ,\quad (L_0+1)(4L_0^2+5L_0+1)>9W_0^2\eqno(A.5) $$
Taking the solutions described by eq.(10-12) we obtain that
they are stable for all possible values of $\Lambda$ and $m$.
In case of the solutions given in eq.(13-15) the second condition
in (A.5) leads to the inequality
$$ (1-y)\bigl[(3x)^2+x( { y \over 2 } +2)+{1\over {(12)^2}}(y-4)^2\bigr]>0
,{\rm \  where }\quad y=\rho ^2 ,\quad x=\Lambda ^2 \eqno(A.6) $$
This implies that the values of $\Lambda$  are not restricted
and the first condition in eq.(A.5) is satisfied as well if $\rho < 1
$.  If  $\rho \geq
1 $ (A.6) does not hold for any value of $\Lambda $ .
This implies, as mentioned in sect.3, that the classical $SL_2$ invariant
vacuum, which
corresponds to $\Lambda=0,\rho=2$, cannot be a classical highest weight state.

\centerline { \bf Appendix B }

\centerline { $W$ matrix elements }

In this appendix we illustrate the method we used to compute the
various matrix elements of $W_{-3}$ on the example
of $W_{-3}u$ between highest weight states:
$$\langle H\Omega \vert W_{-3}u(z)\vert hw\rangle
 =\lim_{z_{1}\to\infty}\lim_{z_{3}\to0}
z_1^{2H}\langle \Phi_{H}(z_{1})W_{-3}u(z)\Phi_{h}(z_{3})\rangle  .\eqno(B.1)$$
(We determined the matrix elements of $L_n$ in the standard way
[20].) In (B.1) $\Phi_{H}(z_1)$and $\Phi_{h}(z_3)$ denote two
(chiral) $W$ primary fields characterized by the $L_0$; $W_0$
eigenvalues $H,\ \Omega $ and $h,\ w$ respectively, generating
the highest weight states from vacuum. We shall use the integral
representation
 $$W_{n}\Phi(z)=\oint \limits_z {d\xi \over 2\pi
i}(\xi-z)^{n+2}W(\xi)u(z) $$
 and the freedom to deform the contour away from $z$ to $z_1$
and $z_3$. For this
 we have to compute
$W(\xi)\Phi_{H}(z_{1})$ and $W(\xi)\Phi_{h}(z_{3})$.

The singular terms in the operator product have the form:
$$W(\xi)\Phi_{H}(z)={\Omega \Phi_{H}(z) \over (\xi-z)^3}
  +{A_{H}(z)\over (\xi-z)^2}+{B_{H}(z)\over (\xi-z)}  \eqno(B.2)$$
where $A_{H}(z)=W_{-1}\Phi_{H}(z)$ and $
  B_{H}(z)=W_{-2}\Phi_{H}(z)$ denote the $W$ descendandts of the
primary field $\Phi_{H}(z)$. It is important to realise that the
irreducible $W$ representation generated from $\Phi_{H}(z)$ may
contain several Virasoro primary fields among the $W$
descendants. If the representation built on $\Phi_{H}(z)$ is not
degenerate on the first grade then the two fields
$L_{-1}\Phi_{H}(z)$ $W_{-1}\Phi_{H}(z)$ are not related to each
other. Therefore defining $\Xi_{H+1}(z)$ as
$$A_{H}(z)={3\Omega \over 2H}L_{-1}\Phi_{H}(z)+\Xi_{H+1}(z) \eqno(B.3)$$
we see using eq.(18) that $\Xi_{H+1}$ is a Virasoro primary field
 $$L_{0}\Xi_{H+1}(z)=(H+1)\Xi_{H+1}(z),\qquad
 L_{n}\Xi_{H+1}(z)=0\quad n>0.$$
In the same way we have
$$B_{H}=AL_{-2}\Phi{H}(z)+BL_{-1}^2\Phi_{H}(z)+DL_{-1}\Xi_{H+1}(z)+
\Psi_{H+2}(z) \eqno(B.4)$$
where $A$, $B$ and $D$ are constants and
 $$L_{0}\Psi_{H+2}(z)=(H+2)\Psi_{H+2},\qquad
L_{n}\Psi_{H+2}(z)=0\quad n>0.$$
(Because $u$ is in an irreducibile  representation degenerate on
the first and second grade
 its associated fields are null, i.e.
$\Xi_{\Delta +1}(z)=0$ and $ \Psi_{\Delta +2}=0$ respectively.)
Since $\Xi_{H+1}(z)$ and $\Psi_{H+2}(z)$ are Virasoro primary
fields, conformal symmetry restricts the $z$ dependence of the
3-point functions they enter.

Deforming the contour in eq.(B.1) to $z_1$ and $z_3$
 and substituting (B.2), (B.3) and (B.4)
 into (B.1) and taking the  $z_{1}\rightarrow \infty $ limit we get:
$$\langle H\Omega \vert W_{-3}u(z) \vert hw\rangle =w{G(H,h,..)\over z^{y+3}}
+{1\over z^2}\langle h\vert u(z)\vert A_{h}\rangle +{1\over z}\langle H\vert
u(z)\vert B_{h}\rangle  \eqno(B.5)$$
where $$\langle H\vert u(z)\vert A_{h}\rangle =\lim_{z_{1}\to
\infty}\lim_{z_{3}\to 0}z_{1}^{2H}\langle \Phi_{H}(z_{1})u(z)
A_{h}(z_{3})\rangle  $$

We shall determine the unknown $\langle H\vert u(z)\vert A_{h}\rangle $
and $\langle H\vert u(z)\vert B_{h}\rangle $ functions by computing
   $\langle H\vert W_{-1}u(z)\vert h\rangle $ and $\langle H\vert
W_{-2}u(z)\vert h\rangle $.
Repeating the same steps that lead from (B.1) to (B.5) we have
$$\langle H\vert W_{-2}u(z)\vert h\rangle =-\langle H\vert
u(z)\vert B_{h}\rangle  \eqno(B.6)$$
and
$$\langle H\vert W_{-1}u(z)\vert h\rangle =z\langle
H\vert u(z)\vert B_{h}\rangle
-\langle H\vert u(z)\vert A_{h}\rangle  .\eqno(B.7)$$
Since the $W$ representation built on $u(z,\bar z)$ is
characterized by the null vectors (21.1), (21.2) in (B.6) and
(B.7) we can write
$$W_{-2}u(z)=\omega \beta^{-1-1}L_{-1}^2u(z)+\omega \beta^{-2}L_{-2}u(z)
\eqno(B.8)$$ and $$W_{-1}u(z)={3\omega \over 2\Delta}L_{-1}u(z) .\eqno(B.9)$$

Substituting (B.6) and (B.7) into (B.5) using (B.8) and (B.9) we get:
$$\langle H\vert W_{-3}u(z)\vert h\rangle ={G(H,h,..)\over z^{y+3}}\bigl\{w+
\omega({3\over 2\Delta}y-2\beta^{-2}(y+h)-2\beta^{-1-1}y(y+1)\bigr\}$$
\vfill
\eject
\centerline{\bf Appendix C}

In this appendix we derive the $x\rightarrow x^{-1}$ transformation
rule for the generalized hypergeometric functions $_3F_2$:
$$_3F_2\Bigl(\matrix{\alpha _1&\alpha _2&\alpha _3\cr
                     \ &\beta _1&\beta _2\cr }\bigl\vert z\Bigr)=
\sum\limits_{n=0}^{\infty }{\alpha _1^{(n)}\alpha _2^{(n)}\alpha _3^{(n)}
\over \beta _1^{(n)}\beta _2^{(n)}}{z^n\over n!}\eqno(C.1)$$
To derive the transformation rule we use an integral representation
wich is a straightforward generalization of the corresponding one for
the hypergeometric functions:
$$\eqalign{{\prod\limits_{i=1}^3\Gamma (\alpha _i)\over
 \prod\limits_{j=1}^2\Gamma (\beta _j)}\
 _3&F_2\Bigl(\matrix{\alpha _1&\alpha _2&\alpha _3\cr
                     \ &\beta _1&\beta _2\cr }\bigl\vert z\Bigr)=\cr
 &={1\over 2\pi i}\int\limits_{-i\infty }^{i\infty }
 {\prod\limits_{i=1}^3\Gamma (\alpha _i+s)\Gamma (-s)\over
 \prod\limits_{j=1}^2\Gamma (\beta _j+s)}(-z)^sds\cr }\eqno(C.2)$$
In (C.2) $\vert {\rm arg}(-z)\vert <\pi $ and the contour of
integration is chosen in such a way that the poles of $\Gamma (\alpha
_i+s)$, $\Gamma (\beta _j+s)$ lie to its left while the poles of
$\Gamma (-s)$ lie to its right. We also assume that none of $\alpha
_i$ is a negative integer. Deforming the contour to encircle the
poles of $\Gamma (-s)$ we indeed recover (C.1). However deforming it
to encircle the poles of $\Gamma (\alpha _i+s)$ we get
$$\eqalign{_3F_2&\Bigl(\matrix{\alpha _1&\alpha _2&\alpha _3\cr
                     \ &\beta _1&\beta _2\cr }\bigl\vert z\Bigr)=\cr &=
\sum\limits_{i=1}^3A_i(-z)^{-\alpha _i}\
_3F_2\Bigl(\matrix{\alpha _i&1+\alpha _i-\beta _1&1+\alpha _i-\beta _2\cr
                     \ &1+\alpha _i-\alpha _{i+1}&1+\alpha _i-\alpha _{i+2}\cr
                      }\bigl\vert z^{-1}\Bigr)\cr }$$
where the $i+1$, $i+2$ indeces are understood only mod 3 and
$$A_i={\Gamma (\beta
_1)\Gamma (\beta _2)\prod\limits_{j\ne i}\Gamma (\alpha _j-\alpha _i)
\over \prod\limits_{j=1}^2\Gamma (\beta _j-\alpha _i)
\prod\limits_{j\ne i}\Gamma (\alpha _j)}\ .$$
\vfill
\eject
\centerline {\bf References}

\item{[1]} J.L. Gervais and A. Neveu : {\sl  Nucl. Phys.} {\bf
B224} (1983) 329.
\item{} E. Braaten, T. Curtright, G. Ghandour and C. Thorn: {\sl Phys.
Lett.} {\bf B125} (1983) 301.
\item{} P. Mansfield : {\sl  Nucl. Phys.} {\bf B222} (1983) 419.
\item{[2]} A. Bilal and J.L. Gervais : {\sl  Nucl. Phys.} {\bf
B318} (1989) 579.
\item{[3]} A.N. Leznov and M.V. Savaliev: {\sl Lett. Math.
Phys.} {\bf 3} (1979) 489; {\sl Comm. Math. Phys.} {\bf 74}
(1980) 111.
\item{[4]} T. Hollowood and P. Mansfield: {\sl Phys.
Lett.} {\bf B226} (1989) 73.
\item{[5]} A. Bilal and J.L. Gervais : {\sl  Phys. Lett.} {\bf
B206} (1988) 412, {\sl Nucl. Phys.}
{\bf B314} (1989) 646.
\item{[6]} A.B. Zamolodchikov : {\sl  Theor. Math. Phys.} {\bf 65}
(1985) 347.
\item{[7]}
P. Forg\'acs, A. Wipf, J. Balog, L. Feh\'er and L. O'Raifeartaigh: {\sl Phys.
Lett.} {\bf B227} (1989) 214; {\bf B244} (1990) 435; {\sl Ann.
Phys. (N. Y.)} {\bf 203} (1990) 76.
\item{[8]} J. Balog, L. Palla: {\sl Phys. Lett.} {\bf B274} (1992) 323.
\item{[9]} V.A. Fateev, A.B. Zamolodchikov: {\sl Nucl. Phys.} {\bf
B280} (1987) 644.
\item{[10]} L. Palla: {\sl Nucl. Phys.} {\bf B341} (1990) 714.
\item{[11]} I.M. Gelfand, L.A. Dikii: {\sl Funk. Anal. Priloz.} {\bf
10} (1976) 13.
\item{[12]} V. Yu. Ovsienko: {\sl Mat. Zam.} {\bf 47} (1990) 62.
\item{} V.Yu. Ovsienko, B.A. Khesin: {\sl Funk. Anal. Priloz.} {\bf
24} (1990) 38.
\item{[13]}
G. P. Jeorjadze, A. K. Pogrebkov and M. C. Polivanov, {\sl
Teor. Mat. Fiz.} {\bf 40} (1979) 221 (in Russian).
\item{[14]} J. Balog, L. Feher, L. Palla unpublished
\item{[15]} V. Knizhnik, A.B. Zamolodchikov: {\sl Nucl. Phys.} {\bf
B247} (1984) 83.
\item{[16]} S. Mizoguchi: {\sl Phys. Lett.} {\bf B222} (1989) 226.
\item{} G.M. Watts: {\sl  Nucl. Phys.} {\bf B326} (1989) 648.
\item{[17]} E.L. Ince Ordinary differential equations Dover Publications, 1956
\item{[18]} J. Vermaseren: FORM User's guide, Nikef Amsterdam, April 1990.
\item{[19]} H. Bateman, A. Erd\'elyi: Higher transcendental functions;
MC Graw Hill Book Company, 1953
\item{[20]} A.A. Belavin, A.M. Polyakov and A.B. Zamolodchikov:{\sl
Nucl. Phys.} {\bf B241} (1984) 333.

\vfill
\end